\documentclass[twocolumn, times]{aastex631}
\usepackage{xspace}
\usepackage{amsmath}
\usepackage{amssymb}
\usepackage{bm}
\usepackage{hyperref}

\newcommand{\edits}[1]{\textcolor{black}{#1}}

\newcommand{\tess}{\emph{TESS}\xspace}

\newcommand{\gaia}{\emph{Gaia}\xspace}

\newcommand{\neid}{NEID\xspace}
\newcommand{\alopeke}{\`{}Alopeke\xspace}
\newcommand{\ut}{UT\xspace}

\newcommand{\tic}{TIC-229742722\xspace}

\newcommand{\toi}{TOI-1859\xspace}
\newcommand{\toib}{TOI-1859b\xspace}

\providecommand{\msun}{\ensuremath{M_\Sun}}
\providecommand{\rsun}{\ensuremath{R_\Sun}}

\providecommand{\rj}{\ensuremath{R_{\rm Jup}}}

\newcommand{\mstar}{$1.287\pm0.061$}
\newcommand{\rstar}{$1.36^{+0.10}_{-0.12}$}
\newcommand{\rhostar}{$0.73^{+0.21}_{-0.15}$}
\newcommand{\logg}{$4.285^{+0.069}_{-0.067}$}
\newcommand{\teff}{$6341^{+68}_{-70}$}
\newcommand{\feh}{$0.120^{+0.054}_{-0.056}$}
\newcommand{\age}{$9.15^{+0.27}_{-0.68}$}
\newcommand{\Gmag}{$10.2883\pm0.0044$}
\newcommand{\GBPmag}{$10.5379\pm0.0028$}
\newcommand{\GRPmag}{$9.8867\pm0.0038$}

\newcommand{\per}{$63.48347\pm0.00010$}
\newcommand{\midt}{$1662.6558\pm0.0014$}

\newcommand{\TESSrprs}{0.0651$^{+0.0010}_{-0.0008}$}
\newcommand{\TESSap}{0.335$^{+0.041}_{-0.040}$}
\newcommand{\TESSrhostar}{0.71$^{+0.17}_{-0.18}$}
\newcommand{\TESSincl}{88.83$^{+0.46}_{-0.58}$}
\newcommand{\TESSecc}{0.66$^{+0.13}_{-0.16}$}
\newcommand{\TESSb}{0.48$^{+0.13}_{-0.09}$}
\newcommand{\TESSrp}{0.86$^{+0.08}_{-0.06}$}

\newcommand{\TESSomega}{77.2$^{+78.1}_{-56.8}$}
\newcommand{\TESSaor}{53.4$^{+4.8}_{-4.0}$}

\newcommand{\DTrprs}{0.0660$^{+0.0004}_{-0.0005}$}
\newcommand{\DTap}{0.337$^{+0.044}_{-0.036}$}
\newcommand{\DTrhostar}{0.73$^{+0.20}_{-0.16}$}
\newcommand{\DTincl}{88.64$^{+0.38}_{-0.28}$}
\newcommand{\DTecc}{0.57$^{+0.12}_{-0.16}$}
\newcommand{\DTvsini}{12.49$^{+0.33}_{-0.33}$}
\newcommand{\DTb}{0.61$^{+0.03}_{-0.03}$}
\newcommand{\DTrp}{0.87$^{+0.07}_{-0.07}$}

\newcommand{\DTnonrotv}{2.87$^{+0.14}_{-0.13}$}
\newcommand{\DTomega}{79.9$^{+83.2}_{-48.2}$}
\newcommand{\DTlam}{38.9$^{+2.8}_{-2.7}$}
\newcommand{\DTaor}{53.7$^{+4.7}_{-4.0}$}

\newcommand{\SERVALrprs}{0.0654$^{+0.0006}_{-0.0006}$}
\newcommand{\SERVALap}{0.336$^{+0.040}_{-0.041}$}
\newcommand{\SERVALrhostar}{0.72$^{+0.17}_{-0.18}$}
\newcommand{\SERVALincl}{88.70$^{+0.43}_{-0.31}$}
\newcommand{\SERVALecc}{0.64$^{+0.11}_{-0.17}$}
\newcommand{\SERVALvsini}{13.27$^{+0.51}_{-0.42}$}
\newcommand{\SERVALb}{0.51$^{+0.07}_{-0.06}$}
\newcommand{\SERVALrp}{0.86$^{+0.07}_{-0.07}$}
\newcommand{\SERVALjitter}{1.51$^{+0.95}_{-1.50}$}
\newcommand{\SERVALomega}{80.5$^{+90.3}_{-45.8}$}
\newcommand{\SERVALlam}{47.5$^{+5.7}_{-6.0}$}
\newcommand{\SERVALaor}{53.5$^{+4.8}_{-4.1}$}

\newcommand{\DRPrprs}{0.0654$^{+0.0007}_{-0.0005}$}
\newcommand{\DRPap}{0.336$^{+0.041}_{-0.038}$}
\newcommand{\DRPrhostar}{0.72$^{+0.19}_{-0.17}$}
\newcommand{\DRPincl}{88.71$^{+0.44}_{-0.31}$}
\newcommand{\DRPecc}{0.64$^{+0.12}_{-0.17}$}
\newcommand{\DRPvsini}{13.22$^{+0.47}_{-0.49}$}
\newcommand{\DRPb}{0.51$^{+0.08}_{-0.05}$}
\newcommand{\DRPrp}{0.87$^{+0.07}_{-0.06}$}
\newcommand{\DRPjitter}{1.52$^{+0.96}_{-1.50}$}
\newcommand{\DRPomega}{78.2$^{+82.6}_{-54.3}$}
\newcommand{\DRPlam}{46.7$^{+7.4}_{-7.6}$}
\newcommand{\DRPaor}{53.5$^{+5.1}_{-4.0}$}

\received{April 3, 2023}
\revised{May 17, 2023}
\accepted{May 25, 2023}

\submitjournal{the AAS Journals}

\shorttitle{A 64-Day Warm Jupiter on an Eccentric and Misaligned Orbit}
\shortauthors{Dong, Wang, Rice, et al.}

\graphicspath{{./}{Figures/}}

\begin{document}

\title{TOI-1859b: A 64-Day Warm Jupiter on an Eccentric and Misaligned Orbit}

\correspondingauthor{Jiayin Dong}
\email{jdong@flatironinstitute.org}

\newcommand{\PSUAA}{Department of Astronomy \& Astrophysics, 525 Davey Laboratory, The Pennsylvania State University, University Park, PA, 16802, USA}
\newcommand{\PSUCEHW}{Center for Exoplanets and Habitable Worlds, 525 Davey Laboratory, The Pennsylvania State University, University Park, PA, 16802, USA}
\newcommand{\PSUICS}{Institute for Computational and Data Sciences, The Pennsylvania State University, University Park, PA, 16802, USA}
\newcommand{\PSUCASt}{Center for Astrostatistics, 525 Davey Laboratory, The Pennsylvania State University, University Park, PA, 16802, USA}
\newcommand{\PSUSETI}{Penn State Extraterrestrial Intelligence Center, 525 Davey Laboratory, The Pennsylvania State University, University Park, PA, 16802, USA}
\newcommand{\FlatironCCA}{Center for Computational Astrophysics, Flatiron Institute, 162 Fifth Avenue, New York, NY 10010, USA}
\newcommand{\Indiana}{Department of Astronomy, Indiana University, Bloomington, IN 47405, USA}
\newcommand{\Yale}{Department of Astronomy, Yale University, New Haven, CT 06511, USA}
\newcommand{\MITKavli}{Department of Physics and Kavli Institute for Astrophysics and Space Research, Massachusetts Institute of Technology, Cambridge, MA 02139, USA}
\newcommand{\MITEaps}{Department of Earth, Atmospheric, and Planetary Sciences, Massachusetts Institute of Technology, Cambridge, MA 02139, USA}
\newcommand{\MITAero}{Department of Aeronautics and Astronautics, Massachusetts Institute of Technology, Cambridge, MA 02139, USA}
\newcommand{\USQ}{University of Southern Queensland, Centre for Astrophysics, West Street, Toowoomba, QLD 4350 Australia}
\newcommand{\CfA}{Center for Astrophysics \textbar \ Harvard \& Smithsonian, 60 Garden Street, Cambridge, MA 02138, USA}
\newcommand{\MSUAstro}{Department of Physics and Astronomy, Michigan State University, East Lansing, MI 48824, USA}
\newcommand{\Birmingham}{School of Physics \& Astronomy, University of Birmingham, Edgbaston, Birmingham B15 2TT, United Kingdom}
\newcommand{\UA}{Steward Observatory, The University of Arizona, 933 N.\ Cherry Ave, Tucson, AZ 85721, USA}
\newcommand{\UAA}{Department of Astronomy and Steward Observatory, University of Arizona, Tucson, AZ 85721, USA}
\newcommand{\Penn}{Department of Physics and Astronomy, University of Pennsylvania, 209 S 33rd St, Philadelphia, PA 19104, USA}
\newcommand{\Caltech}{Department of Astronomy, California Institute of Technology, Pasadena, CA 91125, USA}
\newcommand{\STScI}{Space Telescope Science Institute, 3700 San Martin Dr, Baltimore, MD 21218, USA}
\newcommand{\JHU}{Department of Physics and Astronomy, Johns Hopkins University, 3400 N Charles St, Baltimore, MD 21218, USA}
\newcommand{\GoddardESAL}{Exoplanets and Stellar Astrophysics Laboratory, NASA Goddard Space Flight Center, Greenbelt, MD 20771, USA}
\newcommand{\GoddardISTD}{Instrument Systems and Technology Division, NASA Goddard Space Flight Center, Greenbelt, MD 20771, USA}
\newcommand{\GSFC}{NASA Goddard Space Flight Center, Greenbelt, MD 20771, USA}
\newcommand{\NOIRLab}{NSF's National Optical-Infrared Astronomy Research Laboratory, 950 N.\ Cherry Ave., Tucson, AZ 85719, USA}
\newcommand{\MacquarieSchool}{School of Mathematical and Physical Sciences, Macquarie University, Balaclava Road, North Ryde, NSW 2109, Australia}
\newcommand{\MacquarieCentre}{The Macquarie University Astrophysics and Space Technologies Research Centre, Macquarie University, Balaclava Road, North Ryde, NSW 2109, Australia}
\newcommand{\NIST}{National Institute of Standards \& Technology, 325 Broadway, Boulder, CO 80305, USA}
\newcommand{\CUBoulder}{Department of Physics, 390 UCB, University of Colorado, Boulder, CO 80309, USA}
\newcommand{\JPL}{Jet Propulsion Laboratory, California Institute of Technology, 4800 Oak Grove Drive, Pasadena, California 91109}
\newcommand{\UCI}{Department of Physics \& Astronomy, The University of California, Irvine, Irvine, CA 92697, USA}
\newcommand{\Carleton}{Carleton College, One North College St., Northfield, MN 55057, USA}
\newcommand{\Princeton}{Department of Astrophysical Sciences, Princeton University, 4 Ivy Lane, Princeton, NJ 08540, USA}
\newcommand{\IAS}{Institute for Advance Study, 1 Einstein Drive, Princeton, NJ 08540, USA}
\newcommand{\Tsinghua}{Department of Astronomy, Tsinghua University, Beijing 100084, China}
\newcommand{\Lafayette}{Department of Physics, Lafayette College, 730 High St., Easton, PA 18042, USA}
\newcommand{\NASAAmes}{NASA Ames Research Center, Moffett Field, CA 94035, USA}
\newcommand{\SAI}{Sternberg Astronomical Institute, M.V. Lomonosov Moscow State University, 13, Universitetskij pr., 119234, Moscow, Russia}
\newcommand{\NExScI}{NASA Exoplanet Science Institute, Caltech/IPAC, Mail Code 100-22, 1200 E. California Blvd., Pasadena, CA 91125, USA}
\newcommand{\Vanderbilt}{Department of Physics and Astronomy, Vanderbilt University, Nashville, TN 37235, USA}
\newcommand{\TexasAustin}{Department of Astronomy, The University of Texas at Austin, Austin, TX 78712, USA}
\newcommand{\Kotizarovci}{Kotizarovci Observatory, Sarsoni 90, 51216 Viskovo, Croatia}
\newcommand{\StephenAustin}{Department of Physics, Engineering and Astronomy, Stephen F. Austin State University, 1936 North St, Nacogdoches, TX 75962, USA}
\newcommand{\Oukaimeden}{Oukaimeden Observatory, High Energy Physics and Astrophysics Laboratory, Cadi Ayyad University, Marrakech, Morocco}
\newcommand{\MauryLewin}{The Maury Lewin Astronomical Observatory, Glendora,California.91741. USA}
\newcommand{\ValenciaAstro}{Departamento de Astronom\'{\i}a y Astrof\'{\i}sica, Universidad de Valencia, E-46100 Burjassot, Valencia, Spain}
\newcommand{\ValenciaObs}{Observatorio Astron\'omico, Universidad de Valencia, E-46980 Paterna, Valencia, Spain}
\newcommand{\Wellesley}{Department of Astronomy, Wellesley College, Wellesley, MA 02481, USA}
\newcommand{\Calou}{Observatori de Ca l'Ou, Carrer de dalt 18, Sant Martí Sesgueioles 08282, Barcelona, Spain}
\newcommand{\IACSpain}{Instituto de Astrof\'isica de Canarias (IAC), E-38205 La Laguna, Tenerife, Spain}
\newcommand{\ULLSpain}{Departamento de Astrof\'isica, Universidad de La Laguna (ULL), E-38206 La Laguna, Tenerife, Spain}
\newcommand{\ARU}{Astrobiology Research Unit, Universit\'e de Li\`ege, 19C All\'ee du 6 Ao\^ut, 4000 Li\`ege, Belgium}
\newcommand{\STAR}{Space sciences, Technologies and Astrophysics Research (STAR) Institute, Universit\'e de Li\`ege, Belgium}
\newcommand{\BerkeleyAstro}{Department of Astronomy, University of California, Berkeley, Berkeley, CA, USA}
\newcommand{\MarylandAstro}{Department of Astronomy, University of Maryland, College Park, College Park, MD, USA}
\newcommand{\Dartmouth}{Department of Physics and Astronomy, Dartmouth College, Hanover, NH 03755, USA}
\newcommand{\CarnegieEPL}{Earth and Planets Laboratory, Carnegie Institution for Science, 5241 Broad Branch Road, NW, Washington, DC 20015, USA}
\newcommand{\UCSC}{Department of Astronomy and Astrophysics, University of California, Santa Cruz, CA 95064, USA}
\newcommand{\SETI}{SETI Institute, Carl Sagan Center, 339 Bernardo Ave, Suite 200, Mountain View, CA 94043, USA}
\newcommand{\ETH}{ETH Zurich, Institute for Particle Physics \& Astrophysics, Zurich, Switzerland}
\newcommand{\UNM}{Department of Physics and Astronomy, University of New Mexico, Albuquerque, NM, USA}
\newcommand{\RIACS}{Research Institute for Advanced Computer Science, Universities Space Research Association, Washington, DC 20024, USA}

\author[0000-0002-3610-6953]{Jiayin Dong}
\altaffiliation{Flatiron Research Fellow}
\affiliation{\FlatironCCA}
\affiliation{\PSUAA}
\affiliation{\PSUCEHW}

\author[0000-0002-7846-6981]{Songhu Wang} 
\affiliation{\Indiana}

\author[0000-0002-7670-670X]{Malena Rice} 
\altaffiliation{51 Pegasi b Fellow}
\affiliation{\MITKavli}
\affiliation{\Yale}

\author[0000-0002-4891-3517]{George Zhou} 
\affiliation{\USQ}

\author[0000-0003-0918-7484]{Chelsea X. Huang} 
\affil{\USQ}

\author[0000-0001-9677-1296]{Rebekah I. Dawson} 
\affiliation{\PSUAA}
\affiliation{\PSUCEHW}

\author[0000-0001-7409-5688]{Gudmundur K. Stefánsson} 
\altaffiliation{NASA Hubble Fellow}
\affil{\Princeton}

\author[0000-0003-1312-9391]{Samuel Halverson} 
\affil{\JPL}

\author[0000-0001-8401-4300]{Shubham Kanodia} 
\altaffiliation{Carnegie EPL Fellow}
\affil{\CarnegieEPL}

\author[0000-0001-9596-7983]{Suvrath Mahadevan} 
\altaffiliation{NEID Principal Investigator}
\affil{\PSUAA}
\affil{\PSUCEHW}
\affil{\ETH}

\author[0000-0003-0241-8956]{Michael W.\ McElwain}
\affil{\GoddardESAL}

\author[0000-0003-0353-9741]{Jaime A. Alvarado-Montes} 
\affil{\MacquarieSchool}
\affil{\MacquarieCentre}

\author[0000-0001-8720-5612]{Joe P.\ Ninan} 
\affil{\PSUAA}
\affil{\PSUCEHW}

\author[0000-0003-0149-9678]{Paul Robertson} 
\altaffiliation{NEID Instrument Team Project Scientist}
\affil{\UCI}

\author[0000-0001-8127-5775]{Arpita Roy} 
\affil{\STScI}
\affil{\JHU}

\author[0000-0002-4046-987X]{Christian Schwab} 
\affil{\MacquarieSchool}
\affil{\MacquarieCentre}

\author[0000-0002-9632-9382]{Sarah E.\ Logsdon} 
\affil{\NOIRLab}

\author[0000-0002-4788-8858]{Ryan C. Terrien}
\affil{\Carleton}

\author[0000-0001-6588-9574]{Karen A.\ Collins} 
\affil{\CfA}

\author{Gregor Srdoc} 
\affil{Kotizarovci Observatory, Sarsoni 90, 51216 Viskovo, Croatia}

\author[0000-0003-3904-6754]{Ramotholo Sefako} 
\affiliation{South African Astronomical Observatory, P.O. Box 9, Observatory, Cape Town 7935, South Africa}

\author{Didier Laloum} 
\affiliation{AAVSO, 185 Alewife Brook Parkway, Suite 410 , Cambridge, MA 02138, USA}

\author[0000-0001-9911-7388]{David W. Latham} 
\affil{\CfA}

\author[0000-0001-6637-5401]{Allyson Bieryla} 
\affil{\CfA}

\author[0000-0002-4297-5506]{Paul A.\ Dalba} 
\altaffiliation{51 Pegasi b Fellow}
\affiliation{\UCSC}
\affiliation{\SETI}

\author[0000-0003-2313-467X]{Diana Dragomir} 
\affiliation{\UNM}

\author[0000-0001-6213-8804]{Steven Villanueva Jr.} 
\altaffiliation{NPP Fellow}
\affiliation{NASA Goddard Space Flight Center, Exoplanets and Stellar Astrophysics Laboratory (Code 667), Greenbelt, MD 20771, USA}

\author[0000-0002-2532-2853]{Steve~B.~Howell} 
\affil{\NASAAmes}

\author[0000-0003-2058-6662]{George R. Ricker} 
\affil{\MITKavli}

\author[0000-0002-6892-6948]{S. Seager} 
\affil{\MITEaps}
\affil{\MITKavli}
\affil{\MITAero}

\author[0000-0002-4265-047X]{Joshua N.\ Winn} 
\affil{\Princeton}

\author[0000-0002-4715-9460]{Jon M. Jenkins} 
\affil{\NASAAmes}

\author[0000-0002-1836-3120]{Avi Shporer} 
\affil{\MITKavli}

\author[0000-0003-2196-6675]{David Rapetti} 
\affil{\NASAAmes}
\affil{\RIACS}

\begin{abstract}
Warm Jupiters are close-in giant planets with relatively large planet-star separations (i.e., $10 < a/R_\star < 100$). Given their weak tidal interactions with their host stars, measurements of stellar obliquity may be used to probe the initial obliquity distribution and dynamical history for close-in gas giants.
Using spectroscopic observations, we confirm the planetary nature of TOI-1859b and determine the stellar obliquity of TOI-1859 to be $\lambda =$ \DTlam $\degr$ relative to its planetary companion using the Rossiter-McLaughlin effect. TOI-1859b is a 64-day warm Jupiter orbiting around a late-F dwarf and has an orbital eccentricity of \DTecc\, inferred purely from transit light curves.
The eccentric and misaligned orbit of TOI-1859b is likely an outcome of dynamical interactions, such as planet-planet scattering and planet-disk resonance crossing.
\end{abstract}


\section{Introduction} \label{sec:Intro}

Warm Jupiters are giant planets with planet-star separations $a/R_\star$ of $\sim$10--100. Because of their wide orbital separations, Warm Jupiters are expected to have weak tidal interactions with their host stars, such that they offer new insights into the origin channels of close-in giant planets.
Tidal dissipation in the planet or in the star is not expected to be efficient enough to significantly modify a Warm Jupiter's orbital eccentricity (unless its eccentricity is extremely high, such that $a(1-e^2)$ is small) and its host star's obliquity \citep[see the review by][and references therein]{Ogilvie14}.
Benefiting from this property, a Warm Jupiter's dynamical features may be used as a probe of its original formation environment and dynamical history.

The eccentricities of known Warm Jupiters are broadly distributed from circular to super eccentric orbits (with an eccentricity of 0.8 or higher), and the distribution could be bimodal \citep{Dong21}.
It is yet unclear which dynamical mechanisms predominantly sculpt the eccentricity distribution of Warm Jupiters.
One or more mechanisms, such as planet-planet scattering \citep{Chatterjee08, Nagasawa08, Nagasawa11, Beauge12}, stellar/planetary Kozai (see von Zeipel-Lidov-Kozai oscillations in \citealt{vonZeipel10, Lidov62, Kozai62} and applications in \citealt{Wu03, Fabrycky07, Naoz16, Vick19}), and other secular interactions \citep{Wu11, Petrovich15} could be responsible for the high-$e$ population.
Either smooth disk migration or in-situ formation origins \citep{Duffell15} could feasibly produce the low-$e$ component of the distribution of Warm Jupiters.

Notably, although these mechanisms can all broadly explain the bulk of observed eccentricities of Warm Jupiters, they would generate different stellar obliquities, reflected through observations as different sky-projected spin-orbit angles of Warm Jupiter systems. 
Secular chaos tends to produce mutual inclinations less than $90\degr$ \citep{Teyssandier19}; stellar Kozai predicts bimodal mutual inclinations concentrated at $40\degr$ and $140\degr$ \citep{Fabrycky07, Anderson16, Vick19}; and planet-planet scattering with no Kozai effect predicts mutual inclinations spread from $0\degr$ to $60\degr$ \citep{Chatterjee08}.
Interpreting Warm Jupiters' stellar obliquities with eccentricities is therefore a possible way to distinguish between different proposed mechanisms.

Measuring the stellar obliquity of Warm Jupiter systems via the Rossiter-McLaughlin effect (RM-effect, \citealt{Rossiter24, McLaughlin24}) is challenging because of the rarity of transit opportunities due to Warm Jupiters' long orbital periods. As presented in the review article by \cite{Albrecht22}, only 25 Warm/Cold Jupiters have stellar obliquity measurements, compared to 105 Hot Jupiters. Fortunately, the \textit{Transiting Exoplanet Survey Satellite} \citep[\tess;][]{Ricker15} is discovering a large sample of transiting Warm Jupiters around bright stars that are suitable for the RM-effect follow-up observations \citep[e.g.,][]{Dong21}.

In this work, we confirm the planetary nature of \toib, a 64-day Warm Jupiter on an eccentric orbit, and measure the stellar obliquity using the NEID spectrograph. This measurement is the sixth result from the Stellar Obliquities in Long-period Exoplanet Systems (SOLES) survey \citep{Rice21SOLESI, Wang22SOLESII, Rice22SOLESIII, Rice23SOLESIV}, which was designed to expand the set of spin-orbit constraints for exoplanets on wide orbits. Stellar obliquity constraints for systems spanning a wide range of orbital eccentricities are particularly important to inform whether trends with properties such as host star temperatures \citep{Winn10, Schlaufman10, Albrecht12, Wang21} and planet mass hold as a function of eccentricity \citep[e.g.][]{Rice22HJ}.

In Section~\ref{sec:Observations}, we describe the detection of the \tess transit signal and validation of the signal from ground-based follow-up observations. In Section~\ref{sec:Models}, we present the modeling of stellar properties and planetary properties, and lastly, in Section~\ref{sec:Discussion}, we summarize our discovery and discuss the interpretation of Warm Jupiters' stellar obliquities.

\section{Observations} \label{sec:Observations}

\begin{figure*}[htb!]
    \centering
    \includegraphics{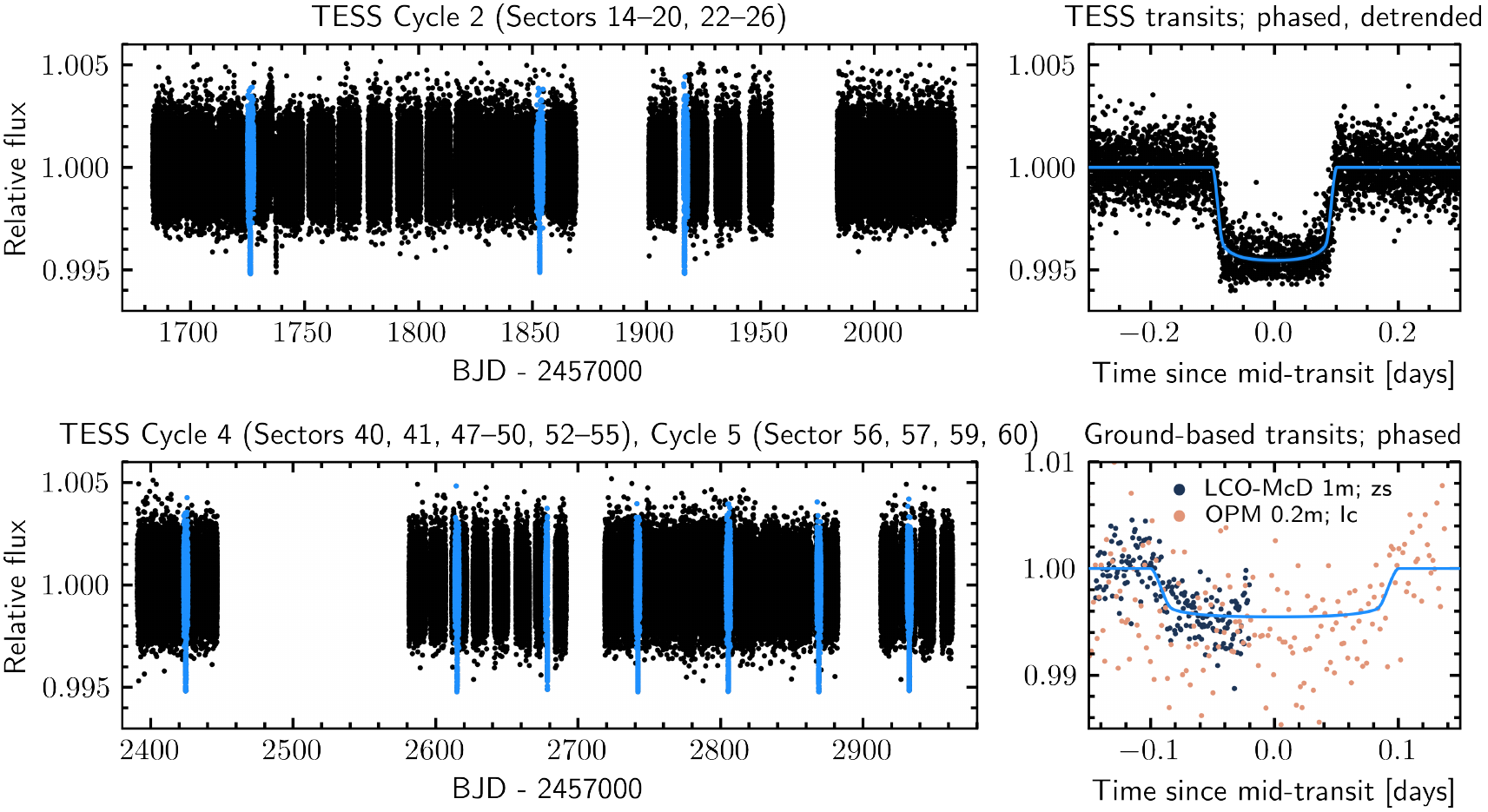}
    \caption{TESS SPOC PDC\_SAP light curves and ground-based transits of \toib. \toib transits are labeled in blue on \tess light curves in the left two panels. The median of the fitted transit model is plotted in blue on the two right panels. Two ground-based transits are shown in the lower right panel.}\label{fig:transit}
\end{figure*}

\subsection{TESS Photometry} \label{subsec:TESS}
As a continuous-viewing-zone target, \toi has been observed in more than 20 \tess sectors over the span of four years. The \tess Cycle 2 observations from Sectors 14--20 and 22--26 (\ut2019-Jul-18 to \ut2020-Jul-04) lead to the initial detection of planetary transit signals. Subsequent Cycle 4 observations in \tess Sectors 40, 41, 47--50, and 52--55 refine the ephemeris and transit timings. In the ongoing \tess Cycle\,5, \toi has been observed for four more sectors from Sectors 56, 57, 59, and 60.
The \tess data products include postage stamps at 2-minute time sampling and full-frame images with 30-minute sampling. Notably, starting from Sector 56, the 20-second cadence postage stamps of \toi are available, enabling future asteroseismology studies.

The transit signal of \toib was detected independently by the NASA Science Processing Operations Center (SPOC) pipeline \citep{Jenkins16} and the MIT Quick-Look Pipeline \citep[QLP;][]{Huang20a, Huang20b}. 
A single transit of TOI-1859b was detected in the SPOC transit search \citep{Jenkins02, Jenkins10, Jenkins20} in Sector 15. The true period was identified in a search of Sector 14--23 in which the signature was fitted with a limb-darkened transit model \citep{Li:DVmodelFit2019} and passed all the Data Validation diagnostic tests \citep{Twicken:DVdiagnostics2018}. A recent search of Sector 14--60 located the source of the transits to within $0.473 \pm 2.6 \arcsec$.
We use the 2-minute cadence SPOC PDC\_SAP light curves \citep{Stumpe2012, Stumpe2014, Smith12}, for which the image data were reduced and analyzed by the SPOC at NASA Ames Research Center, for the analysis in this work. The \tess data are presented in Figure~\ref{fig:transit}.
We found the overall bias on the determination of the planet radius due to bias in the sky background correction in the \tess primary mission (which comprises Sectors 14-26 data) is much smaller than the uncertainty on transit depth, and therefore negligible.

\subsection{Ground-based Transit Photometry} \label{subsec:GB_transits}

Two additional transits of \toib were obtained from ground-based facilities led by the \tess Follow-up Observing Program (TFOP) seeing-limited photometry group \citep{Collins19}. We used the $\mathtt{TESS\,Transit\,Finder}$, which is a customized version of the $\mathtt{Tapir}$ software package \citep{Jensen13}, to schedule our transit observations. Differential photometric data were extracted using $\mathtt{AstroImageJ}$ \citep{Collins17}.
These observations cleared the field for nearby eclipsing binaries and verified that the transit events detected by \tess were on target relative to known Gaia DR3 stars. The transit observations are presented in Figure~\ref{fig:transit}.
\begin{itemize}
    \item We observed a full transit window on \ut2021-03-23 in Ic band from the 0.2\,m telescope at the Private Observatory of the Mount (OPM) near Saint-Pierre-du-Mont, France. The telescope is equipped with a $3326\times2504$ pixel Atik 383 L+ camera having an image scale of $0\farcs69$ per pixel, resulting in a $38\arcmin\times29\arcmin$ field of view. A transit-like event with a depth of $\sim5000$\,ppm was detected using a circular photometric aperture with radius $8\farcs4$ centered on \toi, which excluded most of the flux from the nearest Gaia DR3 neighbor $10\farcs3$ south.
    \item We observed an ingress on \ut2021-05-26 in Pan-STARRS $z$-short band using the Las Cumbres Observatory Global Telescope \citep[LCOGT;][]{Brown:2013} 1.0\,m network node at McDonald Observatory on Mount Fowlkes in Texas, USA. A transit-like ingress with a depth of  $\sim5000$\,ppm was detected using a circular photometric aperture with radius $5\farcs8$ centered on \toi, which excluded flux from the nearest Gaia DR3 neighbor $10\farcs3$ south.
\end{itemize}

\subsection{High-resolution Imaging} \label{subsec:imaging}

\begin{figure}[htb!]
    \centering
    \includegraphics{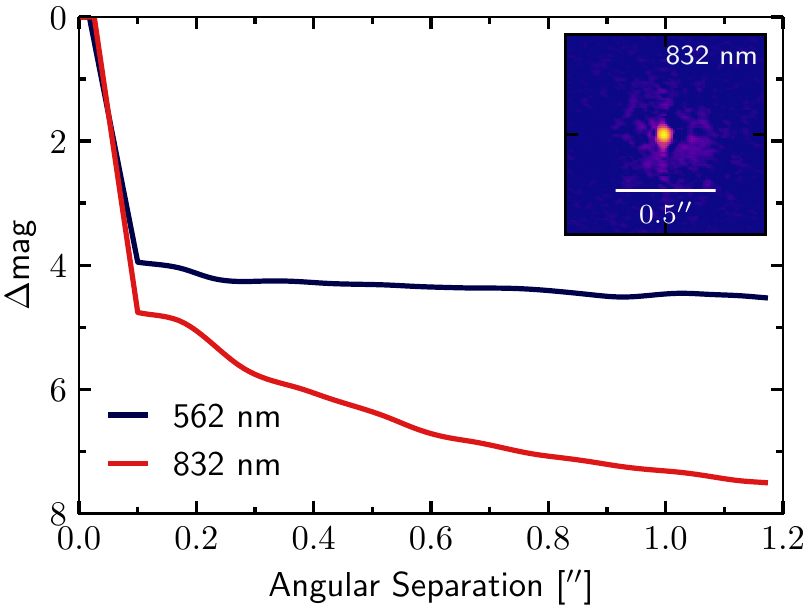}
    \caption{Speckle imaging of \toi taken by the Gemini-North \alopeke. Angular separations from 0.1$\arcsec$ to 1$\arcsec$ correspond to physical distances from 22.4 au to 224 au. Companions are ruled out to a contrast of 6.36 magnitude at $0.5\arcsec$ separation (or a physical distance of 112 au) in the 832\,nm band (red curve and the inset image), and a contrast of 4.31 magnitude at $0.5\arcsec$ separation in the 562\,nm band (blue curve).}\label{fig:imaging}
\end{figure}

High-resolution imaging was performed to search for close companions of \toi.
On \ut2020-06-07, the \alopeke speckle instrument \citep{Scott21} on the 8\,m Gemini North telescope located on Hawaii's Maunakea took simultaneous speckle imaging in 832\,nm and 562\,nm bands. The 5-$\sigma$ contrast curves were achieved. No companion was detected down to a contrast of 6.36 magnitude at $0.5\arcsec$ angular separation in the 832\,nm band, and a contrast of 4.31 magnitude at $0.5\arcsec$ separation in the 562\,nm band, shown in Figure~\ref{fig:imaging}. The angular separation of $0.5\arcsec$ corresponds to a physical distance of 112 au.

\subsection{Long-term Spectroscopic Observation} \label{subsec:recon}
Three spectroscopic observations were taken by Tillinghast Reflector Echelle Spectrograph (TRES) at Whipple Observatory on Mount Hopkins in Arizona, USA.
The TRES spectrograph has a wavelength coverage from 385\,nm to 906\,nm and a resolving power of $R \approx 44,000$ \citep{gaborthesis}. Radial velocities were measured using a cross-correlation analysis against a template spectrum generated from a median combination of all TRES observed spectra \citep{Quinn14}.
Three TRES observations were taken on \ut2020-07-26, \ut2020-09-03, and \ut2022-08-29. We used the highest signal-to-noise spectra taken on \ut2022-08-29 in this study.
The Stellar Parameter Classification code \citep[SPC;][]{Buchhave12, Buchhave14} was applied to extracted TRES spectra \citep{Buchhave10}, and derived stellar parameters are used as priors for the stellar SED joint isochrone modeling in Section~\ref{sec:Models}. 
The SPC derives an effective temperature of $T_\mathrm{eff} = 6318 \pm 100$ K, surface gravity $\log{g} = 4.3 \pm 0.1$ cgs, and bulk metallicity [m/H] $= +0.11 \pm 0.08$ dex. The projected rotational broadening width $v_{\rm broadening} = 13.1 \pm 0.5$ km\,s$^{-1}$, which is introduced by the rotation and macroturbulence of the host star.

Another 23 spectroscopic observations were taken by the Automated Planet Finder (APF) at Lick Observatory on Mount Hamilton in California, USA. The APF spectrograph has a wavelength coverage from 374\,nm to 970\,nm with a maximum resolving power $R \approx 150,000$ \citep{Vogt14, APF}.
The 23 APF observations span over 270 days and present a radial velocity (RV) root-mean-square (RMS) of 59\,m\,s$^{-1}$.  
The scatter in the APF RVs is mainly contributed by the large $v\sin{i_\star}$ of the host star (i.e., $v\sin{i_\star} \sim$ 13\,km\,s$^{-1}$). Also, with $V >$\,10\,mag, it is beyond the limit of APF to yield reliable RVs.

\subsection{Transit Spectroscopy Observation} \label{subsec:NEID}

The transit spectroscopy observation of \toib was taken by the NEID spectrograph \citep{NEID_optical, NEID_budget} on the WIYN 3.5\,m telescope at the Kitt Peak National Observatory (KPNO) in Arizona, USA. The NEID spectrograph is a highly stabilized  \citep{NEID_performance, NEID_stability}, fiber-fed \citep{NEID_fiber} spectrograph with a resolving power of $R\approx110,000$ in high-resolution (HR) mode and a wavelength coverage from 380\,nm to 930\,nm.
The NEID observations were conducted on \ut2022-06-11 from 03:30 to 11:30 for a duration of 8 hours. We obtained 23 spectra, each with a 20-minute exposure, in high-resolution (HR) mode.

The spectra were extracted and radial velocities were reduced by the NEID standard data reduction pipeline $\mathtt{NEID}$-$\mathtt{DRP\,v1.1.4}$ (\url{https://neid.ipac.caltech.edu/docs/NEID-DRP}). 
The $\mathtt{DRP}$ reduces the radial velocities using the cross-correlation of the observed spectra with stellar templates. The barycentric corrected radial velocities for re-weighted orders (CCFRVMOD) were used for the analysis.
We separately reduce the radial velocities using a modified version of the SpEctrum Radial Velocity AnaLyzer ($\mathtt{SERVAL}$) pipeline \cite{zechmeister2018}, which reconstructs the stellar template from observations, optimized for NEID spectra. The NEID customization of the $\mathtt{SERVAL}$ pipeline is discussed in \cite{Stefansson21} Section 3.1. 

The median RV uncertainty from the $\mathtt{DRP}$ pipeline was found to be 10.6,m,s$^{-1}$, while the median RV uncertainty from the $\mathtt{SERVAL}$ pipeline was 6.0,m,s$^{-1}$. 
The RV uncertainty could be overestimated by the $\mathtt{DRP}$ pipeline because of some error in the CCF RV calculation, which could have been further exacerbated by the star's high $v\sin{i_\star}$.
Additionally, the $\mathtt{SERVAL}$ pipeline used more echelle orders than the CCF line mask used in the $\mathtt{DRP}$, which could slightly improve the RV uncertainty. Further investigation is needed to fully understand the factors contributing to the difference in RV uncertainties between the two pipelines.

We also derive line-broadening profiles for each observation to measure the Doppler transit shadow of the planet as a function of velocity and phase. The line profile from each spectrum is derived via a least-squares deconvolution \citep{Donati97} between the \neid spectra and an ATLAS9 synthetic non-rotating spectral template that best fits the host star stellar parameters \citep{Castelli:2004}. See Section 2.6 in \cite{Dong22} for more details.

\begin{figure}
    \centering
    \includegraphics{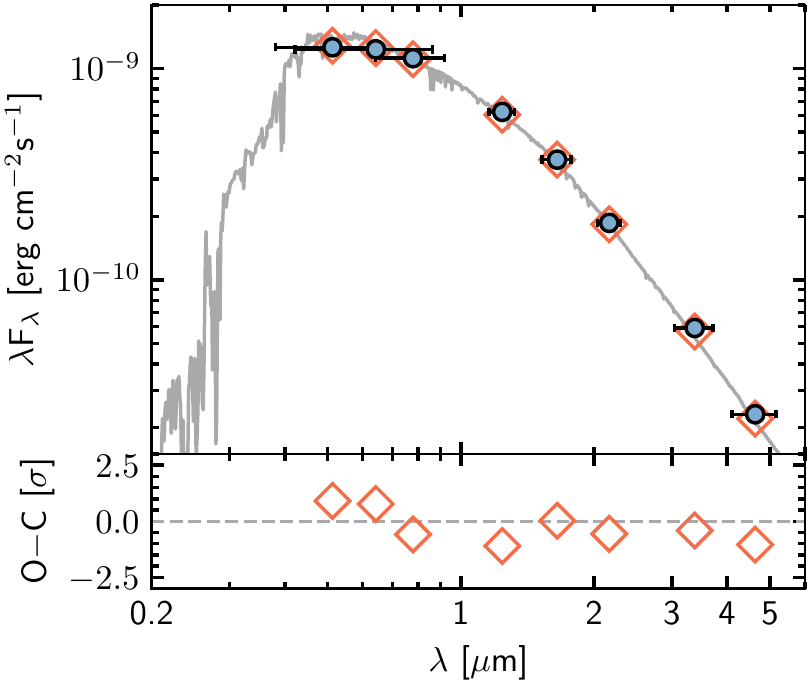}
    \caption{The SED analysis of \toi. Each blue point corresponds to a photometric filter and the x-axis errorbar corresponds to the passband of the filter. The best-fitted SED model is shown in grey. The orange diamond shows the combined synthetic flux over each passband. The SED residuals normalized by the photometric uncertainties are shown in the lower panel.}
    \label{fig:sed}
\end{figure}

\begin{figure*}[htb!]
    \gridline{\fig{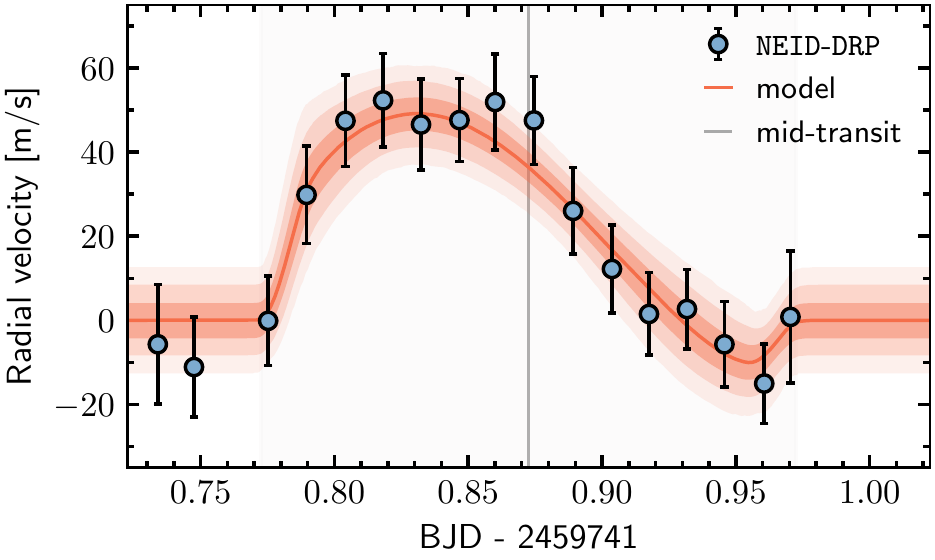}{0.45\textwidth}{\hspace*{1cm}(a) In-transit radial velocities reduced by the $\mathtt{NEID}$-$\mathtt{DRP\,v1.1.4}$ pipeline which infer $\lambda=$\DRPlam$\degr$.}
              \fig{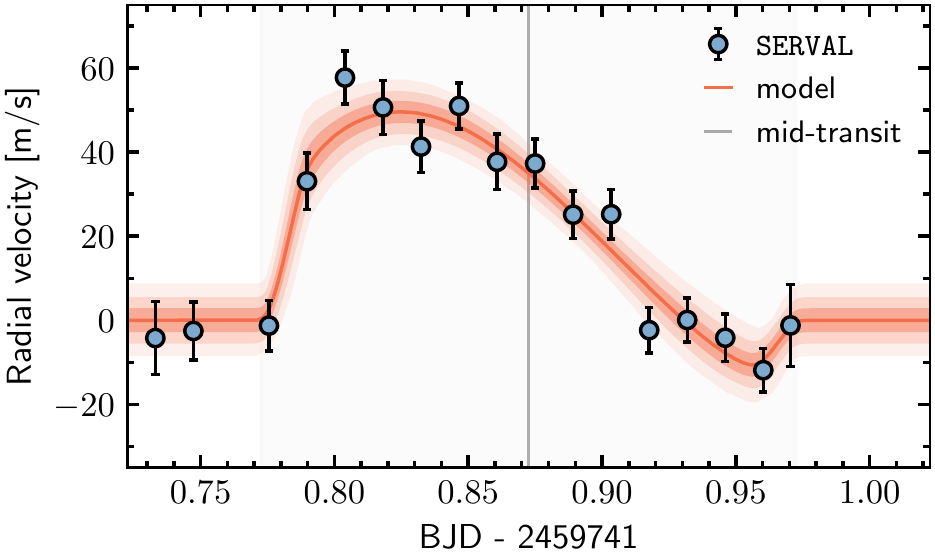}{0.45\textwidth}{\hspace*{1cm}(b) In-transit radial velocities reduced by the $\mathtt{SERVAL}$ pipeline which infer $\lambda=$\SERVALlam$\degr$.}}
    \gridline{\fig{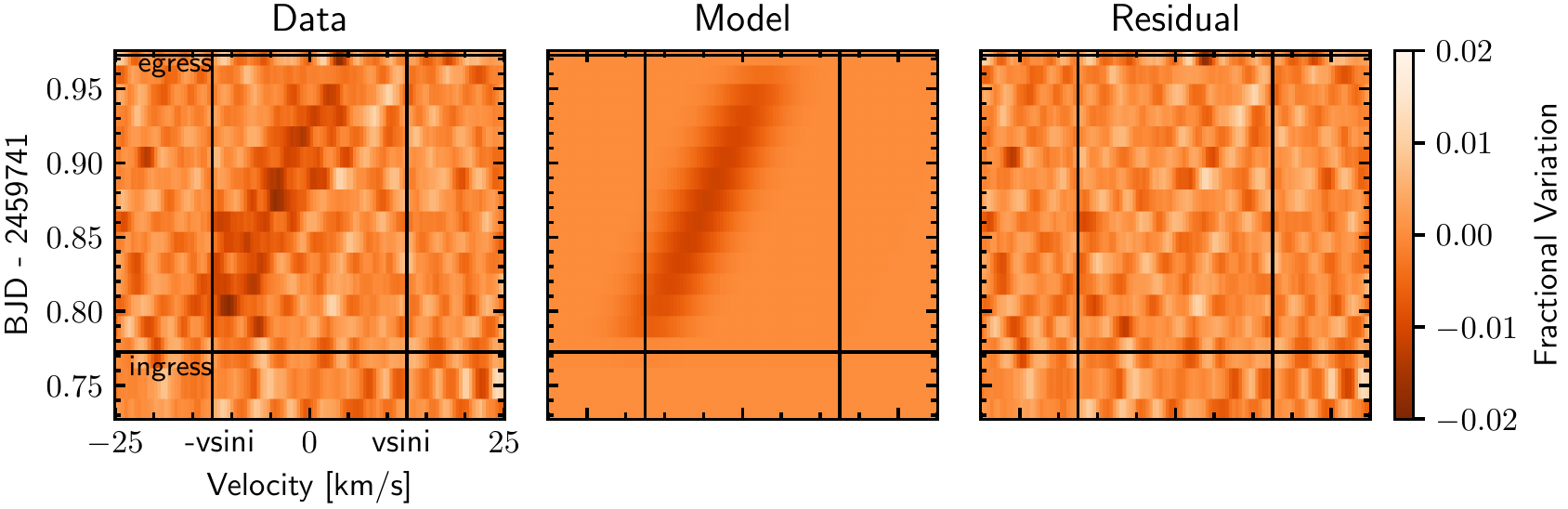}{0.95\textwidth}{(c) Doppler Tomography signal which infers $\lambda=$\DTlam$\degr$.}}
    \caption{\neid spectra are reduced by three different data reduction techniques.
    Inferred projected stellar obliquities are consistent.
    (a) and (b) In-transit radial-velocity measurements of the \toi system using the \neid spectra. The blue dots and black bars are \neid RVs and their corresponding uncertainties. Using the Rossiter-McLaughlin effect, the projected stellar obliquity is constrained. (c) The Doppler Tomography signal of the \toi system during \toib's transit. The left, middle, and right panels are data extracted from the \neid spectra, best-fit model, and the residual of the data after subtracting the best-fit model. The color scale presents the flux variation of the velocity channel.\label{fig:rm}}
\end{figure*}

\begin{figure}
    \centering
    \includegraphics[width=\linewidth]{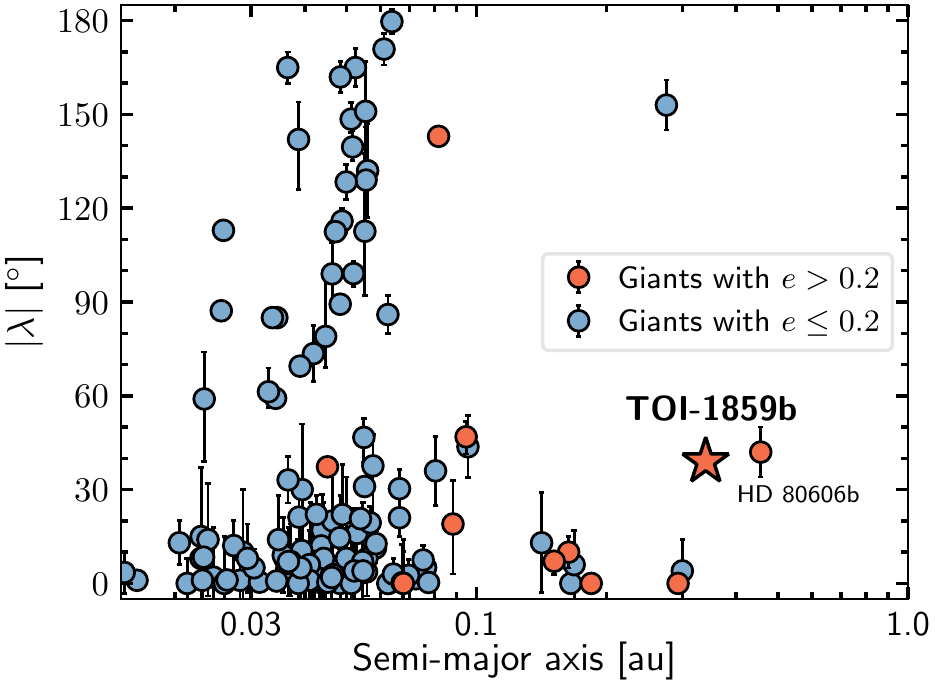}
    \caption{Sky-projected stellar obliquity versus the semi-major axis of all measured giant planets colored by their eccentricities. Eccentric orbits with median $e > 0.2$ are colored in orange and circular orbits with $e \leq 0.2$ are colored in blue. Literature data are extracted from Table A1 in \cite{Albrecht22}.} 
    \label{fig:summary}
\end{figure}

\begin{table*}
\centering
\tabletypesize{\small}
\caption{Median values and 68\% highest density intervals (HDI) for the stellar and planetary parameters of the \toi (\tic) system. \label{tbl:parameters}}
\begin{tabular}{llcccc}
  \hline
  \hline
Parameter & Units & Values \\
\hline\\\multicolumn{2}{l}{Stellar Properties}&\smallskip\\
~~~~$\alpha_{\rm J2016}$\dotfill & \gaia DR3 RA (HH:MM:SS.ss)\dotfill & 18:39:19.62 \\
~~~~$\delta_{\rm J2016}$\dotfill & \gaia DR3 Dec (DD:MM:SS.ss)\dotfill & +69:31:22.34\\
~~~~$\varpi$\dotfill & \gaia DR3 parallax (mas) & $4.4745\pm 0.0123$\\
~~~~$G$\dotfill & \gaia DR3 $G$ magnitude \dotfill & \Gmag\\
~~~~$G_{\mathrm{BP}}$\dotfill & \gaia DR3 $G_{\mathrm{BP}}$ magnitude \dotfill & \GBPmag\\
~~~~$G_{\mathrm{RP}}$\dotfill & \gaia DR3 $G_{\mathrm{RP}}$ magnitude \dotfill & \GRPmag\\
~~~~$M_\star$\dotfill & Stellar mass (\msun)\dotfill & \mstar\\
~~~~$R_\star$\dotfill & Stellar radius (\rsun)\dotfill & \rstar\\
~~~~$\rho_\star$\dotfill & Stellar density (cgs)\dotfill & \rhostar\\
~~~~$\log{g}$\dotfill & Stellar surface gravity (cgs)\dotfill & \logg\\
~~~~$T_{\rm eff}$\dotfill & Stellar effective temperature (K)\dotfill & \teff\\
~~~~$[{\rm m/H}]$\dotfill & Stellar bulk metallicity (dex)\dotfill & \feh\\
~~~~$\log_{10}{\rm Age}$\dotfill & Age (yr)\dotfill & \age\\
~~~~$v\sin i_\star$\dotfill & TRES projected line broadening ($\mathrm{km\,s}^{-1}$)\dotfill & $13.1\pm0.5$\\
\\
\hline\\\multicolumn{2}{l}{Stellar/Planetary Properties from transit$+$RM-effect models}\smallskip\\
~~~~$P$\dotfill & Period (days)\dotfill & \per\\
~~~~$T_C$\dotfill & Mid-transit time (BJD-2457000)\dotfill & \midt\smallskip\\
&& Transit-only & With $\mathtt{NEID}$-$\mathtt{DRP}$ & With $\mathtt{SERVAL}$ & With DT\smallskip\\
~~~~$\rho_\star$ & Fitted stellar density (cgs)\dotfill & \TESSrhostar & \DRPrhostar & \SERVALrhostar & \DTrhostar\\
~~~~$a/R_\star$\dotfill & Planet-star separation\dotfill & \TESSaor & \DRPaor & \SERVALaor & \DTaor\\
~~~~$a$\dotfill & Semi-major axis (au)\dotfill & \TESSap & \DRPap & \SERVALap & \DTap\\
~~~~$b$\dotfill & Impact parameter \dotfill & \TESSb & \DRPb & \SERVALb & \DTb\\
~~~~$i$\dotfill & Orbital inclination ($\degr$)\dotfill & \TESSincl & \DRPincl & \SERVALincl & \DTincl\\
~~~~$R_p/R_\star$\dotfill & Planet-star radius ratio\dotfill & \TESSrprs & \DRPrprs & \SERVALrprs & \DTrprs\\
~~~~$R_p$\dotfill & Planet radius (\rj)\dotfill & \TESSrp & \DRPrp & \SERVALrp & \DTrp\\
~~~~$e$\dotfill & Eccentricity\dotfill & \TESSecc & \DRPecc & \SERVALecc & \DTecc\\
~~~~$\omega$\dotfill & Argument of periapse ($\degr$)\dotfill & \TESSomega & \DRPomega & \SERVALomega & \DTomega\\
~~~~$\lambda$\dotfill & Projected stellar obliquity ($\degr$)\dotfill & - & \DRPlam & \SERVALlam & \DTlam \\
~~~~$v\sin i_\star$\dotfill & Fitted projected line broadening ($\mathrm{km\,s}^{-1}$)\dotfill & - & \DRPvsini & \SERVALvsini & \DTvsini\\
~~~~$v_{\rm macro}$\dotfill & Macroturbulence of the host star ($\mathrm{km\,s}^{-1}$)\dotfill & - & - & - & \DTnonrotv\\
~~~~$\sigma_{\rm RV}$\dotfill & Radial velocity jitter ($\mathrm{m\,s}^{-1}$)\dotfill & - & \DRPjitter & \SERVALjitter & - \\
\smallskip\\
\hline
\end{tabular}
\tablecomments{\gaia magnitudes and spectral line broadening parameter are obtained from the \gaia Data Release 3 \citep{GaiaDR3}. Planetary parameters inferred from the Doppler Tomography signal are used for discussion.}
\end{table*}

\section{Stellar and Planetary Properties} \label{sec:Models}

\subsection{Stellar Modeling}
The spectral energy distribution (SED) of the star is modeled with the $\mathtt{astroARIADNE}$ package \citep{ARIADNE}. 
Our SED analysis of \toi uses magnitudes from eight filters including 2MASS J, H, K, \gaia G, Rp, Bp, and WISE W1, W2. We include an additional error scaling factor for the SED model as recommended in $\mathtt{EXOFASTv2}$ \citep{Eastman:2013, Eastman:2019}. The error scaling factor usually leads to greater but more realistic stellar radius uncertainties. The PHOENIXv2 models \citep{PHOENIXv2} and MIST isochrones \citep{MIST} are applied to the SED and isochrone fits. \edits{The SED and the best-fitted model of \toi are shown in Figure~\ref{fig:sed}.}
For stellar parameter priors, we adopt $T_\mathrm{eff}$ and ${\rm [Fe/H]}$ derived from the TRES spectra.
The inferred stellar parameters are
$M_\star = 1.287\pm0.061\,\msun$, $\log_{10}{\rm Age} = 9.15^{+0.27}_{-0.68}$ (yr), ${\rm [Fe/H]} = 0.12^{+0.054}_{-0.056}$, $R_\star = 1.36^{+0.10}_{-0.12}\,\rsun$, $T_\mathrm{eff} = 6341^{+68}_{-70} {\rm \,K}$, and $A_V = 0.18^{+0.27}_{-0.13}$.
Consequently, we derive $\rho_\star = 0.73^{+0.21}_{-0.15} {\rm \,g\,cm}^{-3}$ and $\log{g} = 4.285^{+0.069}_{-0.067}$.
These stellar parameters are summarized in Table~\ref{tbl:parameters}.

For stellar inclination inference, we looked for periodicities in \tess light curves using the periodogram and ACF methods but detected no clear and consistent rotational modulation signal. Any period in the range from 2--10 days must have a corresponding amplitude smaller than about 0.05\%.

\subsection{Transit Modeling} \label{subsec:transit}

In the transit model, we fit the nine full transits, and one partial transit of \toib detected in \tess SPOC light curves. We model the mid-transit time $T_{0..9}$ for each transit to allow transit-timing variation analysis. \tess light curves are modeled with the $\mathtt{exoplanet}$ package \citep{exoplanet:joss}. We adopt a quadratic limb darkening model and add a Matern-3/2 Gaussian Process (GP) kernel to the transit model.
To reduce computing complexity, \tess light curves are trimmed to roughly 3 times the transit duration before and after the transit.
Even though we model each mid-transit time independently, an orbital period is still required to build the Keplerian orbit of the planet and is fixed at 63.4832 days. The period ($P$) and a reference mid-transit time ($T_C$) are later derived from a linear fit using the posteriors of mid-transit times.
Here we list free parameters in our transit model: $\big\{\rho_\star, b, R_{\rm p}/R_{\star}, T_{0..9}, e, \omega, u_0, u_1\big\}$ and GP hyperparameters $\big\{{\rm GP}_{s, 0..9}, {\rm GP}_{\sigma, 0..9}, {\rm GP}_{\rho, 0..9}\big\}$.
The planet-star radius ratio $R_{\rm p}/R_{\star}$ and GP kernel hyperparameters ${\rm GP}_s, {\rm GP}_\sigma, {\rm GP}_\rho$ are reparameterized to their logarithm forms, eccentricity and argument of periapse $\{e,\omega\}$ to the unit disk distribution as $\{\sqrt{e}\cos{\omega}, \sqrt{e}\sin{\omega}\}$, and limb-darkening coefficients $\{u_0, u_1\}$ to the ones described in \cite{exoplanet:kipping13}.
We take the stellar density $\rho_\star$ prior from the SED fit and use uniform priors for the remaining parameters.
The posteriors are sampled using the gradient-based MCMC algorithm, the No-U-Turn Sampler (NUTS), built in the $\mathtt{PyMC}$ package. A summary of the medians and 94\% highest density intervals (HDI) of the fitted parameters are summarized in Table~\ref{tbl:parameters} in the ``Transit-only" column.

The transit timing variation (TTV) is derived from the deviations of the mid-transit times from a linear trend. \toib's TTV has a root-mean-square (RMS) of 3 minutes, compared to the averaged mid-transit time uncertainty of 2 minutes. The TTV signal does not present an obvious sinusoidal pattern, and it is unclear if the variation is introduced by a nearby companion.

\subsection{Rossiter-McLaughlin Effect Modeling} \label{subsec:rm}

We next model the \neid Rossiter-McLaughlin effect (RM-effect) measurement jointly with the \tess transits. In addition to the previous transit model, we now add another component to fit the radial velocity anomalies due to the Rossiter-McLaughlin effect.
We adopt the expression for the RM-effect radial velocity anomaly in \cite{Hirano11} Equation (16)--(18), which assumes a Gaussian line profile convolving with a rotational-macroturbulent broadening kernel.
To incorporate the integrals in \cite{Hirano11}'s description to the gradient-based sampling code $\mathtt{PyMC}$, we evaluate the integrals numerically using a trapezoidal approximation.
New added parameters include the sky-projected stellar obliquity $\lambda$, the sky-projected rotational broadening of the stellar rotation $v\sin i_\star$, the radial velocity jitter $\sigma_{\rm RV}$, and the quadratic limb-darkening coefficients for the NEID filter. We use a uniform prior on $\lambda$, a Gaussian prior on $v\sin i_\star$ derived from the TRES SPC, and a log uniform prior on $\sigma_{\rm RV}$.
The model also has three more free parameters described in the \cite{Hirano11} model: the Gaussian dispersion of spectral lines $\beta$, the Lorentzian dispersion of spectral lines $\gamma$, and the macroturbulence dispersion $\zeta$. We set uniform priors on these parameters using the typical ranges listed in \cite{Hirano11} Table 1.

For the Doppler Tomography signal, we model the velocity profile of the planetary shadow at each exposure, and from the mean velocity of the planet, we infer the location of the planet on the stellar disk relative to the stellar rotation axis.
For the planetary shadow, we use a Gaussian profile with a width of $\sigma = \sqrt{v_0^2 + v_{\rm macro}^2}$, where $v_0$ is set by the resolution of the spectrograph and $v_{\rm macro}$ is the macroturbulence velocity of the star. A log uniform prior is used on $v_{\rm macro}$, broadly distributed from 0.1--50 km\,s$^{-1}$, to avoid an underestimation on $v_0$.

Again, we use the $\mathtt{PyMC}$ package for the posterior sampling and report the medians and 94\% HDI for the fitted parameters. In Figure~\ref{fig:rm}, we present the RM-effect fit results. The inferred sky-projected stellar obliquity $\lambda$ is \DTlam$\degr$ inferred from the DT signal, \SERVALlam$\degr$ from the $\mathtt{SERVAL}$ RVs, and \DRPlam$\degr$ from the $\mathtt{NEID}$-$\mathtt{DRP}$ RVs.
Planetary parameters inferred from the joint fit are listed in Table~\ref{tbl:parameters}.
Inferred parameters from the $\mathtt{NEID}$-$\mathtt{DRP}$ RVs and the $\mathtt{SERVAL}$ RVs are completely consistent within the uncertainty. The $\lambda$ and $v\sin{i_\star}$ inferred from the RV models are also consistent with the ones inferred from the DT model.

\section{Discussion} \label{sec:Discussion}

\toib is a 64-day, Jupiter-size planet discovered during the \tess mission. 
We confirmed the planetary nature of \toib and measured the obliquity of \toi using the NEID spectrograph. Notably, \toib is the second-longest-period planet with an RM-effect measurement after HD\,80606b (see Figure~\ref{fig:summary}).
We jointly model the planet's transit light curves and the RM-effect signals to constrain its orbital properties. We find a semi-major axis of \DTap\,au, an eccentricity of \DTecc, and a sky-projected stellar obliquity of \DTlam$\degr$. Due to the lack of precise long-term radial velocity data, the eccentricity of \toib is constrained only by the transit light curves.
From the sky-projected stellar obliquity, it is possible to further infer the \emph{true} stellar obliquity of \toi, although the inference will be compromised by the unknown stellar inclination. If we assume an isotropic stellar inclination (i.e., $\cos{i_\star}$ is uniformly distributed), the inferred stellar obliquity is $\psi = 41.5^{+4.9}_{-10.0} \degr$. The mode and 68\% highest density interval are reported since the probability density distribution of $\psi$ is skewed.
\edits{\toib is unlikely to have strong tidal interactions with its host star given its large semi-major axis and moderate eccentricity.}

The speckle imaging rules out nearby companions at the instrumental detection limit. TOI-1859 has a Re-normalised Unit Weight Error (RUWE) number of 1.11, but an excess of astrometric noise.
We performed a cross-match with the Gaia DR3 binaries catalogue \citep{El-Badry21} and identified a likely bound companion at a sky-plane separation of 2360 AU ($\approx 10 \arcsec$) with $G\approx$ 20 mag, with a chance alignment probability of $\approx$ $9\times10^{-4}$ (primary source: Gaia DR3 2259354802894412928; secondary source: Gaia DR3 2259354802893022208). The value of $G_{\rm BP}-G_{\rm RP}$ of the companion is significantly impacted by the contamination of the primary, but from the absolute $G$ magnitude, the companion star is likely a late-type M-dwarf. However, the companion separation is sufficiently large such that it is unlikely to overcome GR precession to induce stellar Kozai oscillation on \toib, when using the criteria in Equation (4) in \cite{Dong14}.

The excitation of \toib's high eccentricity is likely a result of dynamic interactions between multiple planets in the system, such as planet-planet scattering \citep{Nagasawa11, Beauge12} and planetary Kozai \citep{Petrovich16}.
If \toib's eccentricity and inclination are indeed excited by the planetary Kozai, we may expect a nearby planetary companion still coupled to \toib to explain its current relatively low eccentricity. 
The obliquity of \toi could also be consistent with the prediction of the planet-disk resonance crossing due to the protoplanetary disk dissipation in a Warm Jupiter system (\citealt{Petrovich20}; Zanazzi, Chiang in prep). If so, it requires a massive outer companion in the \toi system to allow sufficient angular momentum exchange with \toib.
In any of these scenarios, it is worthwhile to conduct long-term radial velocity follow-up to detect any other planetary/brown dwarf candidates and constrain their mass and orbit.

\toi is a massive, metal-rich star with a stellar mass of \mstar, and a bulk metallicity [m/H] of \feh. \toib's eccentric and misaligned orbit, likely a result of planet-planet interactions, is consistent with the host star metallicity--eccentricity trend found in \cite{Dawson13}. It is also consistent with the ``eccentric migration'' framework proposed by \cite{Wu23}: multiple gas giants are more likely to form around metal-rich stars and can dynamically interact to excite their orbital eccentricities and mutual inclinations.
Furthermore, \toi is a hot star with an effective temperature of \teff\,K. It is yet unclear if the temperature--stellar obliquity trend found in Hot Jupiter systems \citep{Schlaufman10, Winn10} is also true in Warm Jupiter systems. At least for \toib, it is consistent with the trend that misaligned planets are often found around hot stars.
Detecting and characterizing more Warm Jupiter systems like \toib around both cool and hot stars will fill out the parameter space and reveal any possible trends between host star effective temperature and stellar obliquity in Warm Jupiter systems.

\section*{Acknowledgments}
We would like to express our sincere gratitude to Fei Dai, Dan Foreman-Mackey, and Rodrigo Luger for their helpful discussions on data modeling. We would like to thank Kareem El-Badry for his interpretations of the Gaia data.

Data presented were obtained by the NEID spectrograph built by Penn State University and operated at the WIYN Observatory by NOIRLab, under the NN-EXPLORE partnership of the National Aeronautics and Space Administration and the National Science Foundation. These results are based on observations obtained with NEID on the WIYN 3.5m Telescope at Kitt Peak National Observatory (PI Songhu Wang; 2022A-763446 IU TAC). WIYN is a joint facility of the University of Wisconsin–Madison, Indiana University, NSF's NOIRLab, the Pennsylvania State University, Purdue University, University of California, Irvine, and the University of Missouri. The authors are honored to be permitted to conduct astronomical research on Iolkam Du'ag (Kitt Peak), a mountain with particular significance to the Tohono O'odham.

The Pennsylvania State University campuses are located on the original homelands of the Erie, Haudenosaunee (Seneca, Cayuga, Onondaga, Oneida, Mohawk, and Tuscarora), Lenape (Delaware Nation, Delaware Tribe, Stockbridge-Munsee), Shawnee (Absentee, Eastern, and Oklahoma), Susquehannock, and Wahzhazhe (Osage) Nations.  As a land grant institution, we acknowledge and honor the traditional caretakers of these lands and strive to understand and model their responsible stewardship. We also acknowledge the longer history of these lands and our place in that history.
The Center for Exoplanets and Habitable Worlds and the Penn State Extraterrestrial Intelligence Center are supported by the Pennsylvania State University and the Eberly College of Science.

\edits{Some of the data presented in this paper were obtained from the Mikulski Archive for Space Telescopes (MAST) at the Space Telescope Science Institute. The specific observations analyzed can be accessed via \dataset[10.17909/t9-nmc8-f686]{https://doi.org/10.17909/t9-nmc8-f686}.}
This work includes data collected by the TESS mission, which are publicly available from MAST. Funding for the TESS mission is provided by the NASA Science Mission directorate. We acknowledge the use of public TESS data from pipelines at the TESS Science Office and at the TESS Science Processing Operations Center. Resources supporting this work were provided by the NASA High-End Computing (HEC) Program through the NASA Advanced Supercomputing (NAS) Division at Ames Research Center for the production of the SPOC data products. This research has made use of the Exoplanet Follow-up Observation Program website, which is operated by the California Institute of Technology, under contract with the National Aeronautics and Space Administration under the Exoplanet Exploration Program. Some of the data presented in this paper were obtained from MAST. Support for MAST for non-HST data is provided by the NASA Office of Space Science via grant NNX09AF08G and by other grants and contracts. This work has made use of data from the European Space Agency (ESA) mission {\it Gaia} (\url{https://www.cosmos.esa.int/gaia}), processed by the {\it Gaia} Data Processing and Analysis Consortium (DPAC, \url{https://www.cosmos.esa.int/web/gaia/dpac/consortium}). Funding for the DPAC has been provided by national institutions, in particular the institutions participating in the {\it Gaia} Multilateral Agreement. This research has made use of the NASA Exoplanet Archive, which is operated by the California Institute of Technology, under contract with the National Aeronautics and Space Administration under the Exoplanet Exploration Program.

Some of the Observations in the paper made use of the High-Resolution Imaging instrument(s) \alopeke. \alopeke was funded by the NASA Exoplanet Exploration Program and built at the NASA Ames Research Center by Steve B. Howell, Nic Scott, Elliott P. Horch, and Emmett Quigley. \alopeke was mounted on the Gemini North telescope of the international Gemini Observatory, a program of NSF’s NOIRLab, which is managed by the Association of Universities for Research in Astronomy (AURA) under a cooperative agreement with the National Science Foundation on behalf of the Gemini Observatory partnership: the National Science Foundation (United States), National Research Council (Canada), Agencia Nacional de Investigaci\'{o}n y Desarrollo (Chile), Ministerio de Ciencia, Tecnolog\'{i}a e Innovaci\'{o}n (Argentina), Minist\'{e}rio da Ci\^{e}ncia, Tecnologia, Inova\c{c}\~{o}es e Comunica\c{c}\~{o}es (Brazil), and Korea Astronomy and Space Science Institute (Republic of Korea).

This work makes use of observations from the LCOGT network. Part of the LCOGT telescope time was granted by NOIRLab through the Mid-Scale Innovations Program (MSIP). MSIP is funded by NSF.

JAA-M is funded by the International Macquarie University Research Excellence Scholarship (`iMQRES’). GS acknowledges support provided by NASA through the NASA Hubble Fellowship grant HST-HF2-51519.001-A awarded by the Space Telescope Science Institute, which is operated by the Association of Universities for Research in Astronomy, Inc., for NASA, under contract NAS5-26555. 
RID gratefully acknowledges support from the Carnegie Institution for Science Tuve Visiting Scientists program.
This research was carried out, in part, at the Jet Propulsion Laboratory, California Institute of Technology, under a contract with the National Aeronautics and Space Administration (80NM0018D0004).
KAC, SNQ, and DWL acknowledge support from the TESS mission via subaward s3449 from MIT.
D. D. acknowledges support from the NASA Exoplanet Research Program grant 18-2XRP18\_2-0136.
M.R., P.D., and S.W. thank the Heising-Simons Foundation for their generous support. 
DR was supported by NASA under award number NNA16BD14C for NASA Academic Mission Services.

This research made use of $\mathtt{exoplanet}$ \citep{exoplanet:exoplanet, exoplanet:joss} and its dependencies \citep{exoplanet:agol20, exoplanet:astropy13, exoplanet:astropy18, exoplanet:exoplanet, exoplanet:foremanmackey17, exoplanet:foremanmackey18, exoplanet:kipping13, exoplanet:luger19, exoplanet:pymc3, exoplanet:theano}.

\vspace{5mm}
\facilities{\tess, WIYN (\neid), Gemini-North (\alopeke), \gaia, LCOGT, TRES, APF, Exoplanet Archive}

\software{$\mathtt{ArviZ}$ \citep{arviz_2019}, $\mathtt{astroARIADNE}$, $\mathtt{AstroImageJ}$ \citep{Collins17}, $\mathtt{astropy}$ \citep{exoplanet:astropy13, exoplanet:astropy18}, $\mathtt{celerite2}$ \citep{exoplanet:foremanmackey17, exoplanet:foremanmackey18}, $\mathtt{exoplanet}$ \citep{exoplanet:joss, exoplanet:exoplanet}, $\mathtt{Jupyter}$ \citep{Jupyter}, $\mathtt{Matplotlib}$ \citep{Matplotlib07, Matplotlib16}, $\mathtt{NumPy}$ \citep{NumPy11, NumPy20}, $\mathtt{pandas}$ \citep{mckinney-proc-scipy-2010, reback2020pandas}, $\mathtt{PyMC3}$ \citep{exoplanet:pymc3}, $\mathtt{SciPy}$ \citep{2020SciPy-NMeth}, $\mathtt{starry}$ \citep{exoplanet:luger19}, $\mathtt{Tapir}$ \citep{Jensen13}}

\appendix
\section{Supplementary Figures}
\begin{figure}
    \centering
    \includegraphics{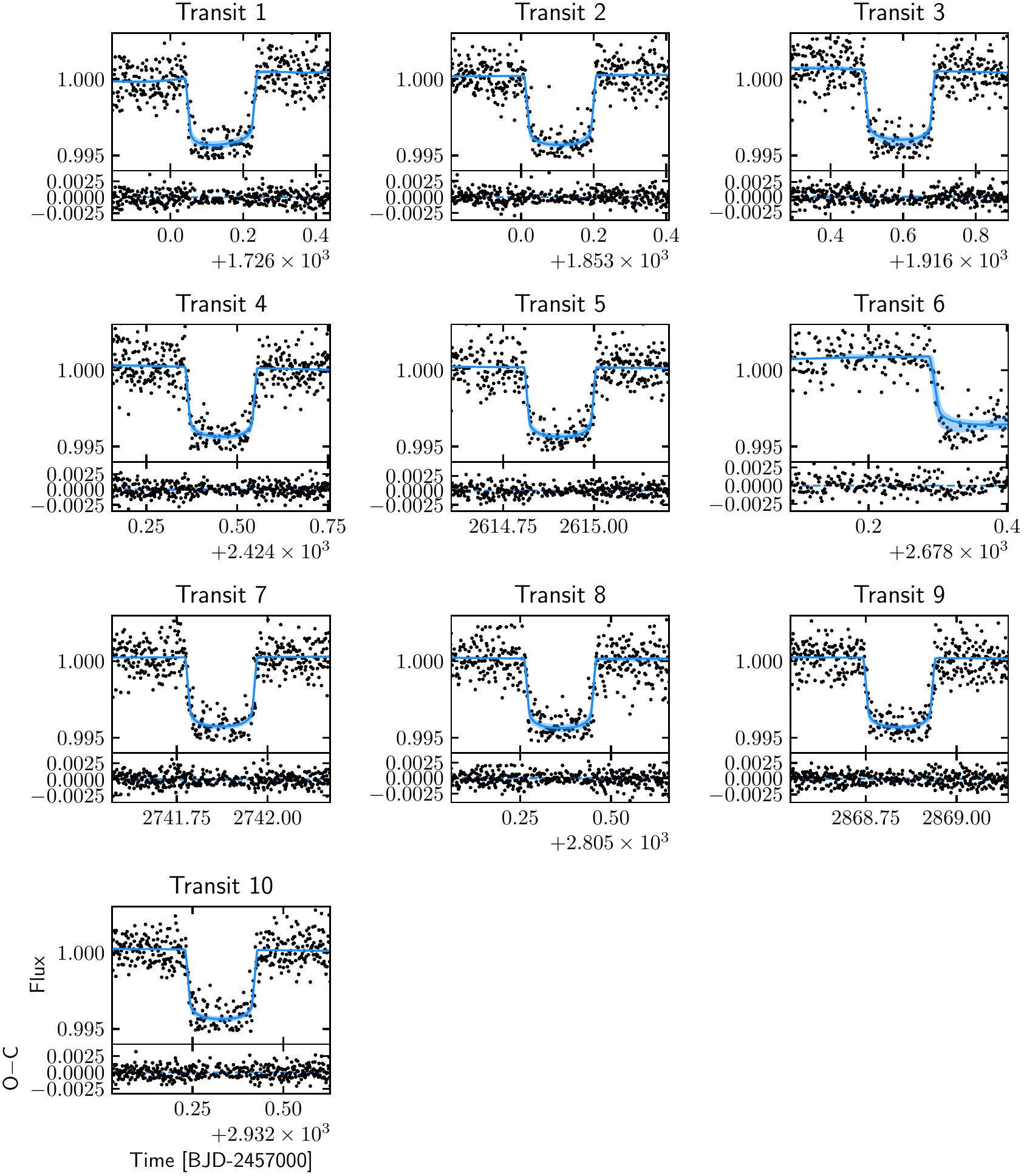}
    \caption{Individual \tess transits of \toib. The medians and 3-$\sigma$ uncertainties of the fitted transit model are plotted in blue lines and  light blue contours, respectively.}
    \label{fig:all_transits}
\end{figure}

\begin{figure}
    \centering
    \includegraphics[width=\linewidth]{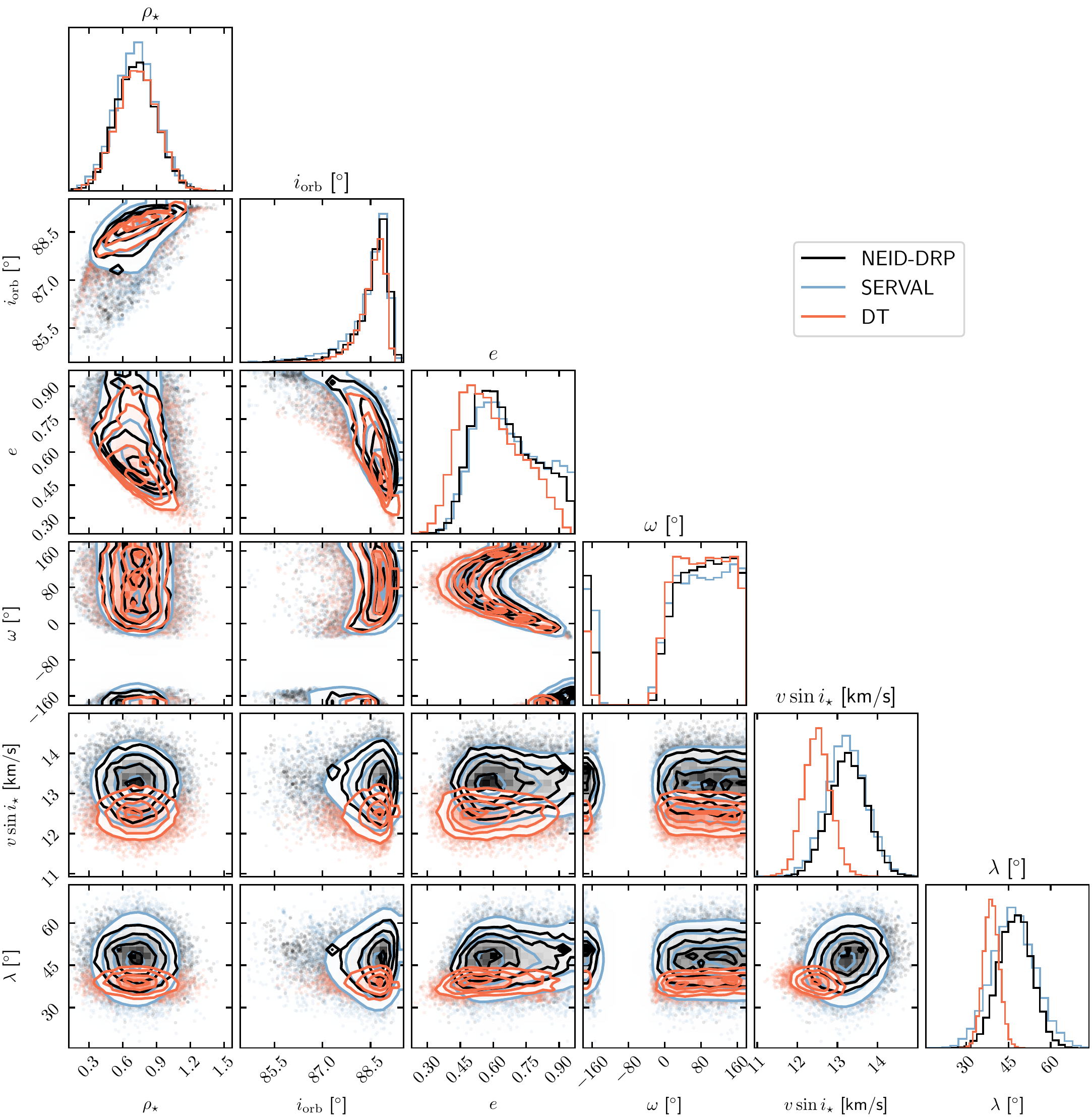}
    \caption{Corner plots of the posteriors of the inferred orbital parameters from the $\mathtt{NEID}$-$\mathtt{DRP}$ radial velocities in black, $\mathtt{SERVAL}$ radial velocities in blue, and Doppler Tomography in orange.}
    \label{fig:corner}
\end{figure}

\edits{
In this section, we include supplementary figures for the stellar and planetary analysis discussed in Section~\ref{sec:Models}.
In Figure~\ref{fig:all_transits}, we show each individual \tess transit light curves of \toib, along with the fitted transit model and residuals.
In Figure~\ref{fig:corner}, we present the corner plot of the posteriors from the transit joint RM-effect fit discussed in Section~\ref{subsec:rm}. Posteriors from the $\mathtt{NEID}$-$\mathtt{DRP}$ radial velocities, $\mathtt{SERVAL}$ radial velocities, and Doppler Tomography data are shown.}

\bibliography{toi1859}{}

\begin{thebibliography}{}
\expandafter\ifx\csname natexlab\endcsname\relax\def\natexlab#1{#1}\fi
\providecommand{\url}[1]{\href{#1}{#1}}
\providecommand{\dodoi}[1]{doi:~\href{http://doi.org/#1}{\nolinkurl{#1}}}
\providecommand{\doeprint}[1]{\href{http://ascl.net/#1}{\nolinkurl{http://ascl.net/#1}}}
\providecommand{\doarXiv}[1]{\href{https://arxiv.org/abs/#1}{\nolinkurl{https://arxiv.org/abs/#1}}}

\bibitem[{{Agol} {et~al.}(2020){Agol}, {Luger}, \&
  {Foreman-Mackey}}]{exoplanet:agol20}
{Agol}, E., {Luger}, R., \& {Foreman-Mackey}, D. 2020, \aj, 159, 123,
  \dodoi{10.3847/1538-3881/ab4fee}

\bibitem[{{Albrecht} {et~al.}(2012){Albrecht}, {Winn}, {Johnson}, {Howard},
  {Marcy}, {Butler}, {Arriagada}, {Crane}, {Shectman}, {Thompson}, {Hirano},
  {Bakos}, \& {Hartman}}]{Albrecht12}
{Albrecht}, S., {Winn}, J.~N., {Johnson}, J.~A., {et~al.} 2012, \apj, 757, 18,
  \dodoi{10.1088/0004-637X/757/1/18}

\bibitem[{{Albrecht} {et~al.}(2022){Albrecht}, {Dawson}, \&
  {Winn}}]{Albrecht22}
{Albrecht}, S.~H., {Dawson}, R.~I., \& {Winn}, J.~N. 2022, \pasp, 134, 082001,
  \dodoi{10.1088/1538-3873/ac6c09}

\bibitem[{{Anderson} {et~al.}(2016){Anderson}, {Storch}, \& {Lai}}]{Anderson16}
{Anderson}, K.~R., {Storch}, N.~I., \& {Lai}, D. 2016, \mnras, 456, 3671,
  \dodoi{10.1093/mnras/stv2906}

\bibitem[{{Astropy Collaboration} {et~al.}(2013){Astropy Collaboration},
  {Robitaille}, {Tollerud}, {Greenfield}, {Droettboom}, {Bray}, {Aldcroft},
  {Davis}, {Ginsburg}, {Price-Whelan}, {Kerzendorf}, {Conley}, {Crighton},
  {Barbary}, {Muna}, {Ferguson}, {Grollier}, {Parikh}, {Nair}, {Unther},
  {Deil}, {Woillez}, {Conseil}, {Kramer}, {Turner}, {Singer}, {Fox}, {Weaver},
  {Zabalza}, {Edwards}, {Azalee Bostroem}, {Burke}, {Casey}, {Crawford},
  {Dencheva}, {Ely}, {Jenness}, {Labrie}, {Lim}, {Pierfederici}, {Pontzen},
  {Ptak}, {Refsdal}, {Servillat}, \& {Streicher}}]{exoplanet:astropy13}
{Astropy Collaboration}, {Robitaille}, T.~P., {Tollerud}, E.~J., {et~al.} 2013,
  \aap, 558, A33, \dodoi{10.1051/0004-6361/201322068}

\bibitem[{{Astropy Collaboration} {et~al.}(2018){Astropy Collaboration},
  {Price-Whelan}, {Sip{\H o}cz}, {G{\"u}nther}, {Lim}, {Crawford}, {Conseil},
  {Shupe}, {Craig}, {Dencheva}, {Ginsburg}, {VanderPlas}, {Bradley},
  {P{\'e}rez-Su{\'a}rez}, {de Val-Borro}, {Aldcroft}, {Cruz}, {Robitaille},
  {Tollerud}, {Ardelean}, {Babej}, {Bach}, {Bachetti}, {Bakanov}, {Bamford},
  {Barentsen}, {Barmby}, {Baumbach}, {Berry}, {Biscani}, {Boquien}, {Bostroem},
  {Bouma}, {Brammer}, {Bray}, {Breytenbach}, {Buddelmeijer}, {Burke},
  {Calderone}, {Cano Rodr{\'{\i}}guez}, {Cara}, {Cardoso}, {Cheedella},
  {Copin}, {Corrales}, {Crichton}, {D'Avella}, {Deil}, {Depagne}, {Dietrich},
  {Donath}, {Droettboom}, {Earl}, {Erben}, {Fabbro}, {Ferreira}, {Finethy},
  {Fox}, {Garrison}, {Gibbons}, {Goldstein}, {Gommers}, {Greco}, {Greenfield},
  {Groener}, {Grollier}, {Hagen}, {Hirst}, {Homeier}, {Horton}, {Hosseinzadeh},
  {Hu}, {Hunkeler}, {Ivezi{\'c}}, {Jain}, {Jenness}, {Kanarek}, {Kendrew},
  {Kern}, {Kerzendorf}, {Khvalko}, {King}, {Kirkby}, {Kulkarni}, {Kumar},
  {Lee}, {Lenz}, {Littlefair}, {Ma}, {Macleod}, {Mastropietro}, {McCully},
  {Montagnac}, {Morris}, {Mueller}, {Mumford}, {Muna}, {Murphy}, {Nelson},
  {Nguyen}, {Ninan}, {N{\"o}the}, {Ogaz}, {Oh}, {Parejko}, {Parley}, {Pascual},
  {Patil}, {Patil}, {Plunkett}, {Prochaska}, {Rastogi}, {Reddy Janga},
  {Sabater}, {Sakurikar}, {Seifert}, {Sherbert}, {Sherwood-Taylor}, {Shih},
  {Sick}, {Silbiger}, {Singanamalla}, {Singer}, {Sladen}, {Sooley},
  {Sornarajah}, {Streicher}, {Teuben}, {Thomas}, {Tremblay}, {Turner},
  {Terr{\'o}n}, {van Kerkwijk}, {de la Vega}, {Watkins}, {Weaver}, {Whitmore},
  {Woillez}, {Zabalza}, \& {Astropy Contributors}}]{exoplanet:astropy18}
{Astropy Collaboration}, {Price-Whelan}, A.~M., {Sip{\H o}cz}, B.~M., {et~al.}
  2018, \aj, 156, 123, \dodoi{10.3847/1538-3881/aabc4f}

\bibitem[{{Beaug{\'e}} \& {Nesvorn{\'y}}(2012)}]{Beauge12}
{Beaug{\'e}}, C., \& {Nesvorn{\'y}}, D. 2012, \apj, 751, 119,
  \dodoi{10.1088/0004-637X/751/2/119}

\bibitem[{{Brown} {et~al.}(2013){Brown}, {Baliber}, {Bianco}, {Bowman},
  {Burleson}, {Conway}, {Crellin}, {Depagne}, {De Vera}, {Dilday}, {Dragomir},
  {Dubberley}, {Eastman}, {Elphick}, {Falarski}, {Foale}, {Ford}, {Fulton},
  {Garza}, {Gomez}, {Graham}, {Greene}, {Haldeman}, {Hawkins}, {Haworth},
  {Haynes}, {Hidas}, {Hjelstrom}, {Howell}, {Hygelund}, {Lister}, {Lobdill},
  {Martinez}, {Mullins}, {Norbury}, {Parrent}, {Paulson}, {Petry}, {Pickles},
  {Posner}, {Rosing}, {Ross}, {Sand}, {Saunders}, {Shobbrook}, {Shporer},
  {Street}, {Thomas}, {Tsapras}, {Tufts}, {Valenti}, {Vander Horst}, {Walker},
  {White}, \& {Willis}}]{Brown:2013}
{Brown}, T.~M., {Baliber}, N., {Bianco}, F.~B., {et~al.} 2013, \pasp, 125,
  1031, \dodoi{10.1086/673168}

\bibitem[{{Buchhave} {et~al.}(2010){Buchhave}, {Bakos}, {Hartman}, {Torres},
  {Kov{\'a}cs}, {Latham}, {Noyes}, {Esquerdo}, {Everett}, {Howard}, {Marcy},
  {Fischer}, {Johnson}, {Andersen}, {F{\H u}r{\'e}sz}, {Perumpilly},
  {Sasselov}, {Stefanik}, {B{\'e}ky}, {L{\'a}z{\'a}r}, {Papp}, \&
  {S{\'a}ri}}]{Buchhave10}
{Buchhave}, L.~A., {Bakos}, G.~{\'A}., {Hartman}, J.~D., {et~al.} 2010, \apj,
  720, 1118, \dodoi{10.1088/0004-637X/720/2/1118}

\bibitem[{{Buchhave} {et~al.}(2012){Buchhave}, {Latham}, {Johansen},
  {Bizzarro}, {Torres}, {Rowe}, {Batalha}, {Borucki}, {Brugamyer}, {Caldwell},
  {Bryson}, {Ciardi}, {Cochran}, {Endl}, {Esquerdo}, {Ford}, {Geary},
  {Gilliland}, {Hansen}, {Isaacson}, {Laird}, {Lucas}, {Marcy}, {Morse},
  {Robertson}, {Shporer}, {Stefanik}, {Still}, \& {Quinn}}]{Buchhave12}
{Buchhave}, L.~A., {Latham}, D.~W., {Johansen}, A., {et~al.} 2012, \nat, 486,
  375, \dodoi{10.1038/nature11121}

\bibitem[{{Buchhave} {et~al.}(2014){Buchhave}, {Bizzarro}, {Latham},
  {Sasselov}, {Cochran}, {Endl}, {Isaacson}, {Juncher}, \&
  {Marcy}}]{Buchhave14}
{Buchhave}, L.~A., {Bizzarro}, M., {Latham}, D.~W., {et~al.} 2014, \nat, 509,
  593, \dodoi{10.1038/nature13254}

\bibitem[{{Castelli} \& {Kurucz}(2004)}]{Castelli:2004}
{Castelli}, F., \& {Kurucz}, R.~L. 2004, ArXiv Astrophysics e-prints

\bibitem[{{Chatterjee} {et~al.}(2008){Chatterjee}, {Ford}, {Matsumura}, \&
  {Rasio}}]{Chatterjee08}
{Chatterjee}, S., {Ford}, E.~B., {Matsumura}, S., \& {Rasio}, F.~A. 2008, \apj,
  686, 580, \dodoi{10.1086/590227}

\bibitem[{{Collins}(2019)}]{Collins19}
{Collins}, K. 2019, in American Astronomical Society Meeting Abstracts, Vol.
  233, American Astronomical Society Meeting Abstracts \#233, 140.05

\bibitem[{{Collins} {et~al.}(2017){Collins}, {Kielkopf}, {Stassun}, \&
  {Hessman}}]{Collins17}
{Collins}, K.~A., {Kielkopf}, J.~F., {Stassun}, K.~G., \& {Hessman}, F.~V.
  2017, \aj, 153, 77, \dodoi{10.3847/1538-3881/153/2/77}

\bibitem[{{Dawson} \& {Murray-Clay}(2013)}]{Dawson13}
{Dawson}, R.~I., \& {Murray-Clay}, R.~A. 2013, \apjl, 767, L24,
  \dodoi{10.1088/2041-8205/767/2/L24}

\bibitem[{{Donati} {et~al.}(1997){Donati}, {Semel}, {Carter}, {Rees}, \&
  {Collier Cameron}}]{Donati97}
{Donati}, J.~F., {Semel}, M., {Carter}, B.~D., {Rees}, D.~E., \& {Collier
  Cameron}, A. 1997, \mnras, 291, 658, \dodoi{10.1093/mnras/291.4.658}

\bibitem[{{Dong} {et~al.}(2021){Dong}, {Huang}, {Dawson}, {Foreman-Mackey},
  {Collins}, {Quinn}, {Lissauer}, {Beatty}, {Quarles}, {Sha}, {Shporer}, {Guo},
  {Kane}, {Abe}, {Barkaoui}, {Benkhaldoun}, {Brahm}, {Bouchy}, {Carmichael},
  {Collins}, {Conti}, {Crouzet}, {Dransfield}, {Evans}, {Gan}, {Ghachoui},
  {Gillon}, {Grieves}, {Guillot}, {Hellier}, {Jehin}, {Jensen}, {Jord{\'a}n},
  {Kamler}, {Kielkopf}, {M{\'e}karnia}, {Nielsen}, {Pozuelos}, {Radford},
  {Schmider}, {Schwarz}, {Stockdale}, {Tan}, {Timmermans}, {Triaud}, {Wang},
  {Ricker}, {Vanderspek}, {Latham}, {Seager}, {Winn}, {Jenkins}, {Mireles},
  {Yahalomi}, {Morgan}, {Vezie}, {Quintana}, {Rose}, {Smith}, \&
  {Shiao}}]{Dong21}
{Dong}, J., {Huang}, C.~X., {Dawson}, R.~I., {et~al.} 2021, \apjs, 255, 6,
  \dodoi{10.3847/1538-4365/abf73c}

\bibitem[{{Dong} {et~al.}(2022){Dong}, {Huang}, {Zhou}, {Dawson},
  {Stef{\'a}nsson}, {Bender}, {Blake}, {Ford}, {Halverson}, {Kanodia},
  {Mahadevan}, {McElwain}, {Ninan}, {Robertson}, {Roy}, {Schwab}, {Stevens},
  {Terrien}, {Vanderburg}, {Kraus}, {Douglas}, {Newton}, {Rampalli},
  {Krolikowski}, {Collins}, {Rodriguez}, {Feliz}, {Srdoc}, {Ziegler},
  {Barkaoui}, {Pozuelos}, {Jehin}, {Micha{\"e}l}, {Benkhaldoun}, {Lewin},
  {For{\'e}s-Toribio}, {Mu{\~n}oz}, {McLeod}, {{\"O}zyurt}, {Horta}, {Murgas},
  {Latham}, {Quinn}, {Bieryla}, {Howell}, {Gnilka}, {Ciardi}, {Lund},
  {Dressing}, {Giacalone}, {Savel}, {Strakhov}, {Belinski}, {Ricker}, {Seager},
  {Winn}, {Jenkins}, {Torres}, \& {Paegert}}]{Dong22}
{Dong}, J., {Huang}, C.~X., {Zhou}, G., {et~al.} 2022, \apjl, 926, L7,
  \dodoi{10.3847/2041-8213/ac4da0}

\bibitem[{{Dong} {et~al.}(2014){Dong}, {Katz}, \& {Socrates}}]{Dong14}
{Dong}, S., {Katz}, B., \& {Socrates}, A. 2014, \apjl, 781, L5,
  \dodoi{10.1088/2041-8205/781/1/L5}

\bibitem[{{Dotter}(2016)}]{MIST}
{Dotter}, A. 2016, \apjs, 222, 8, \dodoi{10.3847/0067-0049/222/1/8}

\bibitem[{{Droettboom} {et~al.}(2016){Droettboom}, {Hunter}, {Caswell},
  {Firing}, {Nielsen}, {Elson}, {Root}, {Dale}, {Lee}, {Sepp{\"a}nen},
  {McDougall}, {Straw}, {May}, {Varoquaux}, {Yu}, {Ma}, {Moad}, {Silvester},
  {Gohlke}, {W{\"u}rtz}, {Hisch}, {Ariza}, {Cimarron}, {Thomas}, {Evans},
  {Ivanov}, {Whitaker}, {Hobson}, {mdehoon}, \& {Giuca}}]{Matplotlib16}
{Droettboom}, M., {Hunter}, J., {Caswell}, T.~A., {et~al.} 2016, {Matplotlib:
  Matplotlib V1.5.1}, v1.5.1,  Zenodo, \dodoi{10.5281/zenodo.44579}

\bibitem[{{Duffell} \& {Chiang}(2015)}]{Duffell15}
{Duffell}, P.~C., \& {Chiang}, E. 2015, \apj, 812, 94,
  \dodoi{10.1088/0004-637X/812/2/94}

\bibitem[{{Eastman} {et~al.}(2013){Eastman}, {Gaudi}, \& {Agol}}]{Eastman:2013}
{Eastman}, J., {Gaudi}, B.~S., \& {Agol}, E. 2013, \pasp, 125, 83,
  \dodoi{10.1086/669497}

\bibitem[{{Eastman} {et~al.}(2019){Eastman}, {Rodriguez}, {Agol}, {Stassun},
  {Beatty}, {Vanderburg}, {Gaudi}, {Collins}, \& {Luger}}]{Eastman:2019}
{Eastman}, J.~D., {Rodriguez}, J.~E., {Agol}, E., {et~al.} 2019, arXiv
  e-prints, arXiv:1907.09480.
\newblock \doarXiv{1907.09480}

\bibitem[{{El-Badry} {et~al.}(2021){El-Badry}, {Rix}, \& {Heintz}}]{El-Badry21}
{El-Badry}, K., {Rix}, H.-W., \& {Heintz}, T.~M. 2021, \mnras, 506, 2269,
  \dodoi{10.1093/mnras/stab323}

\bibitem[{{Fabrycky} \& {Tremaine}(2007)}]{Fabrycky07}
{Fabrycky}, D., \& {Tremaine}, S. 2007, \apj, 669, 1298, \dodoi{10.1086/521702}

\bibitem[{F\H{u}r\'esz(2008)}]{gaborthesis}
F\H{u}r\'esz, G. 2008, PhD thesis, University of Szeged, Hungary

\bibitem[{{Foreman-Mackey}(2018)}]{exoplanet:foremanmackey18}
{Foreman-Mackey}, D. 2018, Research Notes of the American Astronomical Society,
  2, 31, \dodoi{10.3847/2515-5172/aaaf6c}

\bibitem[{{Foreman-Mackey} {et~al.}(2017){Foreman-Mackey}, {Agol},
  {Ambikasaran}, \& {Angus}}]{exoplanet:foremanmackey17}
{Foreman-Mackey}, D., {Agol}, E., {Ambikasaran}, S., \& {Angus}, R. 2017, \aj,
  154, 220, \dodoi{10.3847/1538-3881/aa9332}

\bibitem[{Foreman-Mackey {et~al.}(2019)Foreman-Mackey, Czekala, Luger, Agol,
  Barentsen, \& Barclay}]{exoplanet:exoplanet}
Foreman-Mackey, D., Czekala, I., Luger, R., {et~al.} 2019, dfm/exoplanet:
  exoplanet v0.2.1, \dodoi{10.5281/zenodo.3462740}

\bibitem[{{Foreman-Mackey} {et~al.}(2021){Foreman-Mackey}, {Luger}, {Agol},
  {Barclay}, {Bouma}, {Brandt}, {Czekala}, {David}, {Dong}, {Gilbert},
  {Gordon}, {Hedges}, {Hey}, {Morris}, {Price-Whelan}, \&
  {Savel}}]{exoplanet:joss}
{Foreman-Mackey}, D., {Luger}, R., {Agol}, E., {et~al.} 2021, arXiv e-prints,
  arXiv:2105.01994.
\newblock \doarXiv{2105.01994}

\bibitem[{{Gaia Collaboration} {et~al.}(2022){Gaia Collaboration}, {Vallenari},
  {Brown}, {Prusti}, {de Bruijne}, {Arenou}, {Babusiaux}, {Biermann},
  {Creevey}, {Ducourant}, \& et~al.}]{GaiaDR3}
{Gaia Collaboration}, {Vallenari}, A., {Brown}, A.~G.~A., {et~al.} 2022, arXiv
  e-prints, arXiv:2208.00211.
\newblock \doarXiv{2208.00211}

\bibitem[{{Halverson} {et~al.}(2016){Halverson}, {Terrien}, {Mahadevan}, {Roy},
  {Bender}, {Stef{\'a}nsson}, {Monson}, {Levi}, {Hearty}, {Blake}, {McElwain},
  {Schwab}, {Ramsey}, {Wright}, {Wang}, {Gong}, \& {Roberston}}]{NEID_budget}
{Halverson}, S., {Terrien}, R., {Mahadevan}, S., {et~al.} 2016, in \procspie,
  Vol. 9908, Ground-based and Airborne Instrumentation for Astronomy VI,
  99086P, \dodoi{10.1117/12.2232761}

\bibitem[{{Harris} {et~al.}(2020){Harris}, {Jarrod Millman}, {van der Walt},
  {Gommers}, {Virtanen}, {Cournapeau}, {Wieser}, {Taylor}, {Berg}, {Smith},
  {Kern}, {Picus}, {Hoyer}, {van Kerkwijk}, {Brett}, {Haldane}, {Fern{\'a}ndez
  del R{\'\i}o}, {Wiebe}, {Peterson}, {G{\'e}rard-Marchant}, {Sheppard},
  {Reddy}, {Weckesser}, {Abbasi}, {Gohlke}, \& {Oliphant}}]{NumPy20}
{Harris}, C.~R., {Jarrod Millman}, K., {van der Walt}, S.~J., {et~al.} 2020,
  arXiv e-prints, arXiv:2006.10256.
\newblock \doarXiv{2006.10256}

\bibitem[{{Hirano} {et~al.}(2011){Hirano}, {Suto}, {Winn}, {Taruya}, {Narita},
  {Albrecht}, \& {Sato}}]{Hirano11}
{Hirano}, T., {Suto}, Y., {Winn}, J.~N., {et~al.} 2011, \apj, 742, 69,
  \dodoi{10.1088/0004-637X/742/2/69}

\bibitem[{{Huang} {et~al.}(2020{\natexlab{a}}){Huang}, {Vanderburg}, {P{\'a}l},
  {Sha}, {Yu}, {Fong}, {Fausnaugh}, {Shporer}, {Guerrero}, {Vanderspek}, \&
  {Ricker}}]{Huang20a}
{Huang}, C.~X., {Vanderburg}, A., {P{\'a}l}, A., {et~al.} 2020{\natexlab{a}},
  Research Notes of the American Astronomical Society, 4, 204,
  \dodoi{10.3847/2515-5172/abca2e}

\bibitem[{{Huang} {et~al.}(2020{\natexlab{b}}){Huang}, {Vanderburg}, {P{\'a}l},
  {Sha}, {Yu}, {Fong}, {Fausnaugh}, {Shporer}, {Guerrero}, {Vanderspek}, \&
  {Ricker}}]{Huang20b}
---. 2020{\natexlab{b}}, Research Notes of the American Astronomical Society,
  4, 206, \dodoi{10.3847/2515-5172/abca2d}

\bibitem[{{Hunter}(2007)}]{Matplotlib07}
{Hunter}, J.~D. 2007, Computing in Science and Engineering, 9, 90,
  \dodoi{10.1109/MCSE.2007.55}

\bibitem[{Husser {et~al.}(2013)Husser, von Berg, Dreizler, Homeier, Reiners,
  Barman, \& Hauschildt}]{PHOENIXv2}
Husser, T.-O., von Berg, S.~W., Dreizler, S., {et~al.} 2013, Astronomy {\&}
  Astrophysics, 553, A6, \dodoi{10.1051/0004-6361/201219058}

\bibitem[{{Jenkins}(2002)}]{Jenkins02}
{Jenkins}, J.~M. 2002, \apj, 575, 493, \dodoi{10.1086/341136}

\bibitem[{{Jenkins} {et~al.}(2020){Jenkins}, {Tenenbaum}, {Seader}, {Burke},
  {McCauliff}, {Smith}, {Twicken}, \& {Chandrasekaran}}]{Jenkins20}
{Jenkins}, J.~M., {Tenenbaum}, P., {Seader}, S., {et~al.} 2020, {Kepler Data
  Processing Handbook: Transiting Planet Search}, Kepler Science Document
  KSCI-19081-003

\bibitem[{{Jenkins} {et~al.}(2010){Jenkins}, {Chandrasekaran}, {McCauliff},
  {Caldwell}, {Tenenbaum}, {Li}, {Klaus}, {Cote}, \& {Middour}}]{Jenkins10}
{Jenkins}, J.~M., {Chandrasekaran}, H., {McCauliff}, S.~D., {et~al.} 2010, in
  Society of Photo-Optical Instrumentation Engineers (SPIE) Conference Series,
  Vol. 7740, Software and Cyberinfrastructure for Astronomy, ed. N.~M.
  {Radziwill} \& A.~{Bridger}, 77400D, \dodoi{10.1117/12.856764}

\bibitem[{{Jenkins} {et~al.}(2016){Jenkins}, {Twicken}, {McCauliff},
  {Campbell}, {Sanderfer}, {Lung}, {Mansouri-Samani}, {Girouard}, {Tenenbaum},
  {Klaus}, {Smith}, {Caldwell}, {Chacon}, {Henze}, {Heiges}, {Latham},
  {Morgan}, {Swade}, {Rinehart}, \& {Vanderspek}}]{Jenkins16}
{Jenkins}, J.~M., {Twicken}, J.~D., {McCauliff}, S., {et~al.} 2016, in
  \procspie, Vol. 9913, Software and Cyberinfrastructure for Astronomy IV,
  99133E, \dodoi{10.1117/12.2233418}

\bibitem[{{Jensen}(2013)}]{Jensen13}
{Jensen}, E. 2013, {Tapir: A web interface for transit/eclipse observability},
  Astrophysics Source Code Library.
\newblock \doeprint{1306.007}

\bibitem[{{Kanodia} {et~al.}(2018){Kanodia}, {Mahadevan}, {Ramsey},
  {Stefansson}, {Monson}, {Hearty}, {Blakeslee}, {Lubar}, {Bender}, {Ninan},
  {Sterner}, {Roy}, {Halverson}, \& {Robertson}}]{NEID_fiber}
{Kanodia}, S., {Mahadevan}, S., {Ramsey}, L.~W., {et~al.} 2018, in Society of
  Photo-Optical Instrumentation Engineers (SPIE) Conference Series, Vol. 10702,
  Ground-based and Airborne Instrumentation for Astronomy VII, ed. C.~J.
  {Evans}, L.~{Simard}, \& H.~{Takami}, 107026Q, \dodoi{10.1117/12.2313491}

\bibitem[{{Kipping}(2013)}]{exoplanet:kipping13}
{Kipping}, D.~M. 2013, \mnras, 435, 2152, \dodoi{10.1093/mnras/stt1435}

\bibitem[{Kluyver {et~al.}(2016)Kluyver, Ragan-Kelley, P{\'e}rez, Granger,
  Bussonnier, Frederic, Kelley, Hamrick, Grout, Corlay, Ivanov, Avila, Abdalla,
  Willing, \& development team}]{Jupyter}
Kluyver, T., Ragan-Kelley, B., P{\'e}rez, F., {et~al.} 2016, in Positioning and
  Power in Academic Publishing: Players, Agents and Agendas, ed. F.~Loizides \&
  B.~Scmidt (Netherlands: IOS Press), 87--90.
\newblock \url{https://eprints.soton.ac.uk/403913/}

\bibitem[{{Kozai}(1962)}]{Kozai62}
{Kozai}, Y. 1962, \aj, 67, 591, \dodoi{10.1086/108790}

\bibitem[{Kumar {et~al.}(2019)Kumar, Carroll, Hartikainen, \&
  Martin}]{arviz_2019}
Kumar, R., Carroll, C., Hartikainen, A., \& Martin, O. 2019, Journal of Open
  Source Software, 4, 1143, \dodoi{10.21105/joss.01143}

\bibitem[{{Li} {et~al.}(2019){Li}, {Tenenbaum}, {Twicken}, {Burke}, {Jenkins},
  {Quintana}, {Rowe}, \& {Seader}}]{Li:DVmodelFit2019}
{Li}, J., {Tenenbaum}, P., {Twicken}, J.~D., {et~al.} 2019, \pasp, 131, 024506,
  \dodoi{10.1088/1538-3873/aaf44d}

\bibitem[{{Lidov}(1962)}]{Lidov62}
{Lidov}, M.~L. 1962, \planss, 9, 719, \dodoi{10.1016/0032-0633(62)90129-0}

\bibitem[{{Luger} {et~al.}(2019){Luger}, {Agol}, {Foreman-Mackey}, {Fleming},
  {Lustig-Yaeger}, \& {Deitrick}}]{exoplanet:luger19}
{Luger}, R., {Agol}, E., {Foreman-Mackey}, D., {et~al.} 2019, \aj, 157, 64,
  \dodoi{10.3847/1538-3881/aae8e5}

\bibitem[{{McLaughlin}(1924)}]{McLaughlin24}
{McLaughlin}, D.~B. 1924, \apj, 60, 22, \dodoi{10.1086/142826}

\bibitem[{{Nagasawa} \& {Ida}(2011)}]{Nagasawa11}
{Nagasawa}, M., \& {Ida}, S. 2011, \apj, 742, 72,
  \dodoi{10.1088/0004-637X/742/2/72}

\bibitem[{{Nagasawa} {et~al.}(2008){Nagasawa}, {Ida}, \& {Bessho}}]{Nagasawa08}
{Nagasawa}, M., {Ida}, S., \& {Bessho}, T. 2008, \apj, 678, 498,
  \dodoi{10.1086/529369}

\bibitem[{{Naoz}(2016)}]{Naoz16}
{Naoz}, S. 2016, \araa, 54, 441, \dodoi{10.1146/annurev-astro-081915-023315}

\bibitem[{{Ogilvie}(2014)}]{Ogilvie14}
{Ogilvie}, G.~I. 2014, \araa, 52, 171,
  \dodoi{10.1146/annurev-astro-081913-035941}

\bibitem[{pandas~development team(2020)}]{reback2020pandas}
pandas~development team, T. 2020, pandas-dev/pandas: Pandas, latest,  Zenodo,
  \dodoi{10.5281/zenodo.3509134}

\bibitem[{{Petrovich}(2015)}]{Petrovich15}
{Petrovich}, C. 2015, \apj, 799, 27, \dodoi{10.1088/0004-637X/799/1/27}

\bibitem[{{Petrovich} {et~al.}(2020){Petrovich}, {Mu{\~n}oz}, {Kratter}, \&
  {Malhotra}}]{Petrovich20}
{Petrovich}, C., {Mu{\~n}oz}, D.~J., {Kratter}, K.~M., \& {Malhotra}, R. 2020,
  \apjl, 902, L5, \dodoi{10.3847/2041-8213/abb952}

\bibitem[{{Petrovich} \& {Tremaine}(2016)}]{Petrovich16}
{Petrovich}, C., \& {Tremaine}, S. 2016, \apj, 829, 132,
  \dodoi{10.3847/0004-637X/829/2/132}

\bibitem[{{Quinn} {et~al.}(2014){Quinn}, {White}, {Latham}, {Buchhave},
  {Torres}, {Stefanik}, {Berlind}, {Bieryla}, {Calkins}, {Esquerdo},
  {F{\H{u}}r{\'e}sz}, {Geary}, \& {Szentgyorgyi}}]{Quinn14}
{Quinn}, S.~N., {White}, R.~J., {Latham}, D.~W., {et~al.} 2014, \apj, 787, 27,
  \dodoi{10.1088/0004-637X/787/1/27}

\bibitem[{{Radovan} {et~al.}(2014){Radovan}, {Lanclos}, {Holden}, {Kibrick},
  {Allen}, {Deich}, {Rivera}, {Burt}, {Fulton}, {Butler}, \& {Vogt}}]{APF}
{Radovan}, M.~V., {Lanclos}, K., {Holden}, B.~P., {et~al.} 2014, in Society of
  Photo-Optical Instrumentation Engineers (SPIE) Conference Series, Vol. 9145,
  Ground-based and Airborne Telescopes V, ed. L.~M. {Stepp}, R.~{Gilmozzi}, \&
  H.~J. {Hall}, 91452B, \dodoi{10.1117/12.2057310}

\bibitem[{{Rice} {et~al.}(2023){Rice}, {Wang}, {Gerbig}, {Wang}, {Dai},
  {Tyler}, {Isaacson}, \& {Howard}}]{Rice23SOLESIV}
{Rice}, M., {Wang}, S., {Gerbig}, K., {et~al.} 2023, \aj, 165, 65,
  \dodoi{10.3847/1538-3881/aca88e}

\bibitem[{{Rice} {et~al.}(2022{\natexlab{a}}){Rice}, {Wang}, \&
  {Laughlin}}]{Rice22HJ}
{Rice}, M., {Wang}, S., \& {Laughlin}, G. 2022{\natexlab{a}}, \apjl, 926, L17,
  \dodoi{10.3847/2041-8213/ac502d}

\bibitem[{{Rice} {et~al.}(2021){Rice}, {Wang}, {Howard}, {Isaacson}, {Dai},
  {Wang}, {Beard}, {Behmard}, {Brinkman}, {Rubenzahl}, \&
  {Laughlin}}]{Rice21SOLESI}
{Rice}, M., {Wang}, S., {Howard}, A.~W., {et~al.} 2021, \aj, 162, 182,
  \dodoi{10.3847/1538-3881/ac1f8f}

\bibitem[{{Rice} {et~al.}(2022{\natexlab{b}}){Rice}, {Wang}, {Wang},
  {Stef{\'a}nsson}, {Isaacson}, {Howard}, {Logsdon}, {Schweiker}, {Dai},
  {Brinkman}, {Giacalone}, \& {Holcomb}}]{Rice22SOLESIII}
{Rice}, M., {Wang}, S., {Wang}, X.-Y., {et~al.} 2022{\natexlab{b}}, \aj, 164,
  104, \dodoi{10.3847/1538-3881/ac8153}

\bibitem[{{Ricker} {et~al.}(2015){Ricker}, {Winn}, {Vanderspek}, {Latham},
  {Bakos}, {Bean}, {Berta-Thompson}, {Brown}, {Buchhave}, {Butler}, {Butler},
  {Chaplin}, {Charbonneau}, {Christensen-Dalsgaard}, {Clampin}, {Deming},
  {Doty}, {De Lee}, {Dressing}, {Dunham}, {Endl}, {Fressin}, {Ge}, {Henning},
  {Holman}, {Howard}, {Ida}, {Jenkins}, {Jernigan}, {Johnson}, {Kaltenegger},
  {Kawai}, {Kjeldsen}, {Laughlin}, {Levine}, {Lin}, {Lissauer}, {MacQueen},
  {Marcy}, {McCullough}, {Morton}, {Narita}, {Paegert}, {Palle}, {Pepe},
  {Pepper}, {Quirrenbach}, {Rinehart}, {Sasselov}, {Sato}, {Seager},
  {Sozzetti}, {Stassun}, {Sullivan}, {Szentgyorgyi}, {Torres}, {Udry}, \&
  {Villasenor}}]{Ricker15}
{Ricker}, G.~R., {Winn}, J.~N., {Vanderspek}, R., {et~al.} 2015, Journal of
  Astronomical Telescopes, Instruments, and Systems, 1, 014003,
  \dodoi{10.1117/1.JATIS.1.1.014003}

\bibitem[{{Robertson} {et~al.}(2019){Robertson}, {Anderson}, {Stefansson},
  {Hearty}, {Monson}, {Mahadevan}, {Blakeslee}, {Bender}, {Ninan}, {Conran},
  {Levi}, {Lubar}, {Cole}, {Dykhouse}, {Kanodia}, {Nitroy}, {Smolsky},
  {Tuggle}, {Blank}, {Nelson}, {Blake}, {Halverson}, {Henderson}, {Kaplan},
  {Li}, {Logsdon}, {McElwain}, {Rajagopal}, {Ramsey}, {Roy}, {Schwab},
  {Terrien}, \& {Wright}}]{NEID_performance}
{Robertson}, P., {Anderson}, T., {Stefansson}, G., {et~al.} 2019, Journal of
  Astronomical Telescopes, Instruments, and Systems, 5, 015003,
  \dodoi{10.1117/1.JATIS.5.1.015003}

\bibitem[{{Rossiter}(1924)}]{Rossiter24}
{Rossiter}, R.~A. 1924, \apj, 60, 15, \dodoi{10.1086/142825}

\bibitem[{Salvatier {et~al.}(2016)Salvatier, Wiecki, \&
  Fonnesbeck}]{exoplanet:pymc3}
Salvatier, J., Wiecki, T.~V., \& Fonnesbeck, C. 2016, PeerJ Computer Science,
  2, e55

\bibitem[{{Schlaufman}(2010)}]{Schlaufman10}
{Schlaufman}, K.~C. 2010, \apj, 719, 602, \dodoi{10.1088/0004-637X/719/1/602}

\bibitem[{{Schwab} {et~al.}(2016){Schwab}, {Rakich}, {Gong}, {Mahadevan},
  {Halverson}, {Roy}, {Terrien}, {Robertson}, {Hearty}, {Levi}, {Monson},
  {Wright}, {McElwain}, {Bender}, {Blake}, {St{\"u}rmer}, {Gurevich},
  {Chakraborty}, \& {Ramsey}}]{NEID_optical}
{Schwab}, C., {Rakich}, A., {Gong}, Q., {et~al.} 2016, in \procspie, Vol. 9908,
  Ground-based and Airborne Instrumentation for Astronomy VI, 99087H,
  \dodoi{10.1117/12.2234411}

\bibitem[{{Scott} {et~al.}(2021){Scott}, {Howell}, {Gnilka}, {Stephens},
  {Salinas}, {Matson}, {Furlan}, {Horch}, {Everett}, {Ciardi}, {Mills}, \&
  {Quigley}}]{Scott21}
{Scott}, N.~J., {Howell}, S.~B., {Gnilka}, C.~L., {et~al.} 2021, Frontiers in
  Astronomy and Space Sciences, 8, 138, \dodoi{10.3389/fspas.2021.716560}

\bibitem[{{Smith} {et~al.}(2012){Smith}, {Stumpe}, {Van Cleve}, {Jenkins},
  {Barclay}, {Fanelli}, {Girouard}, {Kolodziejczak}, {McCauliff}, {Morris}, \&
  {Twicken}}]{Smith12}
{Smith}, J.~C., {Stumpe}, M.~C., {Van Cleve}, J.~E., {et~al.} 2012, \pasp, 124,
  1000, \dodoi{10.1086/667697}

\bibitem[{{Stefansson} {et~al.}(2016){Stefansson}, {Hearty}, {Robertson},
  {Mahadevan}, {Anderson}, {Levi}, {Bender}, {Nelson}, {Monson}, {Blank},
  {Halverson}, {Henderson}, {Ramsey}, {Roy}, {Schwab}, \&
  {Terrien}}]{NEID_stability}
{Stefansson}, G., {Hearty}, F., {Robertson}, P., {et~al.} 2016, \apj, 833, 175,
  \dodoi{10.3847/1538-4357/833/2/175}

\bibitem[{{Stefansson} {et~al.}(2021){Stefansson}, {Mahadevan}, {Petrovich},
  {Winn}, {Kanodia}, {Maney}, {Ca{\~n}as}, {Wisniewski}, {Robertson}, {Ninan},
  {Ford}, {Bender}, {Blake}, {Cegla}, {Cochran}, {Diddams}, {Dong}, {Endl},
  {Fredrick}, {Halverson}, {Hearty}, {Hebb}, {Hirano}, {Lin}, {Logsdon},
  {Lubar}, {McElwain}, {Metcalf}, {Monson}, {Rajagopal}, {Ramsey}, {Roy},
  {Schwab}, {Schweiker}, {Terrien}, \& {Wright}}]{Stefansson21}
{Stefansson}, G., {Mahadevan}, S., {Petrovich}, C., {et~al.} 2021, arXiv
  e-prints, arXiv:2111.01295.
\newblock \doarXiv{2111.01295}

\bibitem[{{Stumpe} {et~al.}(2014){Stumpe}, {Smith}, {Catanzarite}, {Van Cleve},
  {Jenkins}, {Twicken}, \& {Girouard}}]{Stumpe2014}
{Stumpe}, M.~C., {Smith}, J.~C., {Catanzarite}, J.~H., {et~al.} 2014, \pasp,
  126, 100, \dodoi{10.1086/674989}

\bibitem[{{Stumpe} {et~al.}(2012){Stumpe}, {Smith}, {Van Cleve}, {Twicken},
  {Barclay}, {Fanelli}, {Girouard}, {Jenkins}, {Kolodziejczak}, {McCauliff}, \&
  {Morris}}]{Stumpe2012}
{Stumpe}, M.~C., {Smith}, J.~C., {Van Cleve}, J.~E., {et~al.} 2012, \pasp, 124,
  985, \dodoi{10.1086/667698}

\bibitem[{{Teyssandier} {et~al.}(2019){Teyssandier}, {Lai}, \&
  {Vick}}]{Teyssandier19}
{Teyssandier}, J., {Lai}, D., \& {Vick}, M. 2019, \mnras, 486, 2265,
  \dodoi{10.1093/mnras/stz1011}

\bibitem[{{Theano Development Team}(2016)}]{exoplanet:theano}
{Theano Development Team}. 2016, arXiv e-prints, abs/1605.02688.
\newblock \url{http://arxiv.org/abs/1605.02688}

\bibitem[{{Twicken} {et~al.}(2018){Twicken}, {Catanzarite}, {Clarke},
  {Girouard}, {Jenkins}, {Klaus}, {Li}, {McCauliff}, {Seader}, {Tenenbaum},
  {Wohler}, {Bryson}, {Burke}, {Caldwell}, {Haas}, {Henze}, \&
  {Sanderfer}}]{Twicken:DVdiagnostics2018}
{Twicken}, J.~D., {Catanzarite}, J.~H., {Clarke}, B.~D., {et~al.} 2018, \pasp,
  130, 064502, \dodoi{10.1088/1538-3873/aab694}

\bibitem[{{van der Walt} {et~al.}(2011){van der Walt}, {Colbert}, \&
  {Varoquaux}}]{NumPy11}
{van der Walt}, S., {Colbert}, S.~C., \& {Varoquaux}, G. 2011, Computing in
  Science and Engineering, 13, 22, \dodoi{10.1109/MCSE.2011.37}

\bibitem[{{Vick} {et~al.}(2019){Vick}, {Lai}, \& {Anderson}}]{Vick19}
{Vick}, M., {Lai}, D., \& {Anderson}, K.~R. 2019, \mnras, 484, 5645,
  \dodoi{10.1093/mnras/stz354}

\bibitem[{{Vines} \& {Jenkins}(2022)}]{ARIADNE}
{Vines}, J.~I., \& {Jenkins}, J.~S. 2022, \mnras, \dodoi{10.1093/mnras/stac956}

\bibitem[{Virtanen {et~al.}(2020)Virtanen, Gommers, Oliphant, Haberland, Reddy,
  Cournapeau, Burovski, Peterson, Weckesser, Bright, {van der Walt}, Brett,
  Wilson, Millman, Mayorov, Nelson, Jones, Kern, Larson, Carey, Polat, Feng,
  Moore, {VanderPlas}, Laxalde, Perktold, Cimrman, Henriksen, Quintero, Harris,
  Archibald, Ribeiro, Pedregosa, {van Mulbregt}, \& {SciPy 1.0
  Contributors}}]{2020SciPy-NMeth}
Virtanen, P., Gommers, R., Oliphant, T.~E., {et~al.} 2020, Nature Methods, 17,
  261, \dodoi{10.1038/s41592-019-0686-2}

\bibitem[{{Vogt} {et~al.}(2014){Vogt}, {Radovan}, {Kibrick}, {Butler},
  {Alcott}, {Allen}, {Arriagada}, {Bolte}, {Burt}, {Cabak}, {Chloros},
  {Cowley}, {Deich}, {Dupraw}, {Earthman}, {Epps}, {Faber}, {Fischer}, {Gates},
  {Hilyard}, {Holden}, {Johnston}, {Keiser}, {Kanto}, {Katsuki}, {Laiterman},
  {Lanclos}, {Laughlin}, {Lewis}, {Lockwood}, {Lynam}, {Marcy}, {McLean},
  {Miller}, {Misch}, {Peck}, {Pfister}, {Phillips}, {Rivera}, {Sandford},
  {Saylor}, {Stover}, {Thompson}, {Walp}, {Ward}, {Wareham}, {Wei}, \&
  {Wright}}]{Vogt14}
{Vogt}, S.~S., {Radovan}, M., {Kibrick}, R., {et~al.} 2014, \pasp, 126, 359,
  \dodoi{10.1086/676120}

\bibitem[{{von Zeipel}(1910)}]{vonZeipel10}
{von Zeipel}, H. 1910, Astronomische Nachrichten, 183, 345,
  \dodoi{10.1002/asna.19091832202}

\bibitem[{{Wang} {et~al.}(2021){Wang}, {Winn}, {Addison}, {Dai}, {Rice},
  {Holden}, {Burt}, {Wang}, {Butler}, {Vogt}, \& {Laughlin}}]{Wang21}
{Wang}, S., {Winn}, J.~N., {Addison}, B.~C., {et~al.} 2021, \aj, 162, 50,
  \dodoi{10.3847/1538-3881/ac0626}

\bibitem[{{Wang} {et~al.}(2022){Wang}, {Rice}, {Wang}, {Pu}, {Stef{\'a}nsson},
  {Mahadevan}, {Radzom}, {Giacalone}, {Wu}, {Esposito}, {Dalba}, {Avsar},
  {Holden}, {Skiff}, {Polakis}, {Voeller}, {Logsdon}, {Klusmeyer}, {Schweiker},
  {Wu}, {Beard}, {Dai}, {Lubin}, {Weiss}, {Bender}, {Blake}, {Dressing},
  {Halverson}, {Hearty}, {Howard}, {Huber}, {Isaacson}, {Jackman}, {Llama},
  {McElwain}, {Rajagopal}, {Roy}, {Robertson}, {Schwab}, {Shkolnik}, {Wright},
  \& {Laughlin}}]{Wang22SOLESII}
{Wang}, X.-Y., {Rice}, M., {Wang}, S., {et~al.} 2022, \apjl, 926, L8,
  \dodoi{10.3847/2041-8213/ac4f44}

\bibitem[{{W}es {M}c{K}inney(2010)}]{mckinney-proc-scipy-2010}
{W}es {M}c{K}inney. 2010, in {P}roceedings of the 9th {P}ython in {S}cience
  {C}onference, ed. {S}t\'efan van~der {W}alt \& {J}arrod {M}illman, 56 -- 61,
  \dodoi{10.25080/Majora-92bf1922-00a}

\bibitem[{{Winn} {et~al.}(2010){Winn}, {Fabrycky}, {Albrecht}, \&
  {Johnson}}]{Winn10}
{Winn}, J.~N., {Fabrycky}, D., {Albrecht}, S., \& {Johnson}, J.~A. 2010, \apjl,
  718, L145, \dodoi{10.1088/2041-8205/718/2/L145}

\bibitem[{{Wu} {et~al.}(2023){Wu}, {Rice}, \& {Wang}}]{Wu23}
{Wu}, D.-H., {Rice}, M., \& {Wang}, S. 2023, arXiv e-prints, arXiv:2302.12778,
  \dodoi{10.48550/arXiv.2302.12778}

\bibitem[{{Wu} \& {Lithwick}(2011)}]{Wu11}
{Wu}, Y., \& {Lithwick}, Y. 2011, \apj, 735, 109,
  \dodoi{10.1088/0004-637X/735/2/109}

\bibitem[{{Wu} \& {Murray}(2003)}]{Wu03}
{Wu}, Y., \& {Murray}, N. 2003, \apj, 589, 605, \dodoi{10.1086/374598}

\bibitem[{{Zechmeister} {et~al.}(2018){Zechmeister}, {Reiners}, {Amado},
  {Azzaro}, {Bauer}, {B{\'e}jar}, {Caballero}, {Guenther}, {Hagen}, {Jeffers},
  {Kaminski}, {K{\"u}rster}, {Launhardt}, {Montes}, {Morales}, {Quirrenbach},
  {Reffert}, {Ribas}, {Seifert}, {Tal-Or}, \& {Wolthoff}}]{zechmeister2018}
{Zechmeister}, M., {Reiners}, A., {Amado}, P.~J., {et~al.} 2018, \aap, 609,
  A12, \dodoi{10.1051/0004-6361/201731483}

\end{thebibliography}
\bibliographystyle{aasjournal}

\end{document}